\documentclass[journal]{IEEEtran}
\usepackage{setspace}

\usepackage{cite}

\ifCLASSINFOpdf
\else
\fi
\usepackage{amsmath}
\usepackage{algorithmic,algorithm}
\usepackage{amsthm,multirow}
\usepackage{amsmath,amssymb,cleveref}
\newtheorem{theorem}{Theorem}
\newtheorem{lemma}{Lemma}
\newtheorem{corollary}{Corollary}

\usepackage{enumerate}
\usepackage{epsfig}
\usepackage{graphicx}
\usepackage{subcaption}
\usepackage{color}
\usepackage[utf8]{inputenc}
\newtheorem{rem}{Remark}
\usepackage{array}

\usepackage{stfloats}
\begin{document}
%
\title{Modeling and Estimation for Systems with Randomly Delayed Measurements and Packet Dropouts}
%
%
%

\author{Ranjeet~Kumar~Tiwari, and~Shovan~Bhaumik
\thanks{Ranjeet Kumar Tiwari and Shovan Bhaumik are with the Department of Electrical Engineering, Indian Institute of Technology Patna, India (e-mail: ranjeet.pee16@iitp.ac.in and shovan.bhaumik@iitp.ac.in.)}}
\maketitle

\begin{abstract}
 A networked system often uses a shared communication network to transmit the measurements to a remotely located estimation center. Due to the limited bandwidth of the channel, a delay may appear while receiving the measurements. This delay can be arbitrary step random, and packets are sometimes dropped during transmission as it exceeds a certain permissible number. In this paper, such measurements are modeled with the Poisson distribution, which allows the user to determine the maximum delay the system might suffer. When the measurement delay exceeds the permissible number, the packet dropout happens. Based on the proposed model, we solve the problem by assuming that the prior and posterior densities of states are Gaussian and derive the expression of the estimated state and the error covariance. Later, relaxing the Gaussian assumption for densities, we propose a solution with the help of the sequential Monte Carlo (SMC) approach. The proposed SMC method divides the set of particles into several groups, where each group supports the possibility that the received measurement is delayed by a certain number of steps. The strength of an individual group is determined by the probability of a measurement being delayed with the same number of steps that the group represents. This approach estimates the states and also assesses the amount of delay from the received measurements. Finally, the developed estimators are implemented on two nonlinear estimation problems, and the simulation results are compared. The proposed SMC approach shows better results compared to the designed Gaussian delay filters and existing particle filters with delay.
\end{abstract}


%
\IEEEpeerreviewmaketitle

\section{Introduction}
%
%
%
%
\IEEEPARstart{T}{he} networked control systems are widely used in various areas such as unmanned aerial vehicles \cite{khamseh2019unscented}, terrestrial and space exploration \cite{abbondanza2017jtrf2014}, and accessing the hazardous environment \cite{hu2016estimation}, to name a few. In such systems, the information is sent over a common communication channel with limited capacity, which inevitably causes some undesirable events such as random delay in measurements, missing measurements/packet dropouts, sensor saturation, and signal quantization, among other networked-induced phenomena. Consequently, these unwanted events warrant some modification in the conventional algorithms of state estimation for the said system.  In this paper, we have considered the event of random delays in measurements and packet dropouts for developing the Bayesian estimators.  

In literature, several works have addressed the random delay in measurement while designing the Gaussian state estimator \cite{he2013optimal,ma2017distributed,wang2016measurement,yan2017networked}. An unscented Kalman filter using the state augmentation is proposed in \cite{hermoso2009unscented} when the measurements are two-step randomly delayed. Assuming the bounded delayed measurements and packet dropouts, the author in \cite{sun2009linear} has designed an optimal estimator for the networked control system. The same author has proposed an optimal linear filter for delayed measurements with and without time-stamps \cite{sun2012optimal}. A generalized filtering methodology for a Gaussian system with maximum one-step random delay in measurements is presented in \cite{wang2013gaussian} whereas \cite{singh2016modified} has dealt with multiple-step delayed measurements along with packet dropouts while developing a generic Gaussian filter. The nonlinear estimators mentioned above have assumed that the system remains Gaussian even when subjected to nonlinearity in dynamics and random delay in measurements. Moreover, they have used the measurement models where a measurement can be received more than once; consequently, the received measurement does not remain independent of past measurements, and the measurement noise sequence becomes correlated over time.

For addressing the non-Gaussian systems, a particle filtering solution for one-step randomly delayed measurement by modifying the importance weight is proposed in \cite{zhang2016particle}. Later, the same authors extended their work for multiple-step delayed measurements \cite{huang2015particle}. With a different measurement model that incorporates packet dropouts as well as randomly delayed measurements, \cite{tiwari2020particle} has also presented a particle filtering method for estimating the states. These works have used the sum of the product of likelihood densities employing every particle repeatedly at the successive time steps for the entire length of maximum delay. They also used the measurement models that generate the dependent measurements and correlated noise sequence.

 In this technical note, we propose a new measurement model to represent the random delay and packet dropouts in measurements by employing the Poisson random variables. This generalized model can cover a range of delay and packet drop scenarios by just varying a single parameter. It does not allow any measurement to be received more than once; hence, the measurement noise sequences are not correlated. Further, we first derive the generalized Gaussian-approximated estimator for the presented delay model by computing the terms that get affected by the random delay in measurements. 

Then, we propose a sequential Monte Carlo (SMC) method that does not mandatorily seek a Gaussian system for the state estimation in the delayed environment. The proposed SMC algorithm divides the set of particles sampled from the proposal density into various groups, where each group supports the possibility that the received measurement is delayed by a certain number of steps. Subsequently, the importance weight of particles is computed based on the delay group to which they are assigned. Finally, the resampling is carried out to select the particles that effectively support the received measurement. Moreover, in this approach, the delay value assigned to the resampled particles conveys the information on the delay steps of the received measurement, which can be estimated at each step. Unlike the sum of products of the likelihood densities method used in existing particle filtering solutions, the proposed method avoids the repeated use of the same particle in computing the importance weight. Hence, the relevant particles get a higher chance of representing the posterior density of the state.

Two nonlinear state estimation problems have been simulated to validate the proposed Bayesian estimators by comparing their performances with that of the existing estimators. The simulation results demonstrate the effectiveness of the proposed estimators.         
%
%

\section{System Representation with Randomly Delayed Measurements}
Consider a nonlinear dynamic system that can be described by the following equations:
\begin{equation} \label{state_dynamics}
	\textit{State equation:}\qquad x_k=f_{k-1}(x_{k-1})+\eta_{k-1},
\end{equation}
\begin{equation} \label{ideal_meas}
	\textit{Measurement equation:}\qquad z_k=h_k(x_k)+v_k,
\end{equation}
where $ x_k \in \Re^{n_x} $ denotes the state vector of the system and $ z_k \in \Re^{n_z} $ is the measurement at any discrete time $ k \in (0,1,\cdots) $, while $ \eta_{k-1} \in \Re^{n_x} $ and $ v_k \in \Re^{n_z} $ are mutually independent white noises with arbitrary but known probability density function (pdf).
 Consider a case where the received measurement might be a randomly delayed measurement from a previous time step owing to limitations, such as small bandwidth and communication failures, of the communication network inserted between the sensor and the estimator. Assume that the measurements are not time-stamped, and at a given time step, a maximum of one measurement can be received at the estimator end. In literature, many models depict similar scenarios with different properties.
 \subsection{Existing Models for Randomly delayed Measurements}
  In literature, the Bernoulli random variables are mostly used to represent the random delay in measurements. Majority of the existing Bernoulli-based multiple-step delay models can be explained with help of the following two models:
 \begin{enumerate}[(i)]
 	\item The delayed measurement, $y_k$, is expressed as \cite{huang2015particle}
 	\begin{equation}
 		y_k=\sum_{j=0}^N\beta_k^jz_{k-j}; \quad k\geq 2,
 	\end{equation} 
 	where 
 	\begin{equation*}
 		\beta_k^j=\begin{cases}
 			\prod_{i=0}^{j} \alpha_k^i (1-\alpha_k^{j+1});& \ 0\leq j<N, \ \text{and} \ \alpha_k^0=1\\
 			\prod_{i=1}^j \alpha_k^i;& \ j=N.
 		\end{cases}
 	\end{equation*}
 	Here, $\alpha_k^i$ are Bernoulli random numbers with $\mathbb{E}[\alpha_k^i]=\Theta$, and at any instant, only one of $\beta_k^j$ can be $1$ and the rest are zero. A set of received measurements using this model is simulated in Table~\ref{chap2_tab0}, where the second row represents the value of index $j$ when $\alpha_k^j=1$.
 	\begin{table}[h]
 		\caption{Received measurements with $\Theta=0.5$ and $N=2$}
 		\label{chap2_tab0}
 		\begin{center}
 			\begin{tabular}{|c|p{0.2cm}|c|c|c|c|c|c|c|c|c|}
 				\hline
 				$k$ & 1&2&3&4&5&6&7&8&9&10\\
 				\hline
 				$j$&$0$&$1$&$0$&$1$&$0$&$2$&$0$&$1$&$0$&$0$\\
 				\hline
 				$y_k$&$z_1$&$z_1$&$z_3$&$z_3$&$z_5$&$z_4$&$z_7$&$z_7$&$z_9$&$z_{10}$\\
 				\hline	
 			\end{tabular}
 		\end{center}
 	\end{table}
 	The authors in \cite{zhang2016particle, wang2013gaussian, zhou2008h, wang2014design} have used the Bernoulli distribution to model one step random delay, and \cite{hermoso2009unscented}  modeled for two steps delay with the help of it. One step or two steps delay model are basically a special case of the above model of multi-step delay when $N$ is set to $1$ and $2$, respectively. In these models, the measurement is mandatorily received at each time step and packet drops are not allowed to occur.
 	\item The received measurement, $y_k$, is represented as \cite{singh2016modified}
 	\begin{equation}
 		y_k=\sum_{j=0}^N\beta_k^jz_{k-j} + \left(1-\sum_{j=0}^N\beta_k^j\right)y_{k-1}; \qquad k\geq 2,
 	\end{equation}
 	where $\beta_k^j=\prod_{i=0}^j\alpha_k^i(1-\alpha_k^{j+1}).$ Further, $\alpha_k^i$ are the Bernoulli random variables with $\mathbb{E}[\alpha_k^i]=\Theta$, and $\beta_k^j$  are the binary variables, which, at a given time instant $k$,  can be $1$ for only one of $j$ $(0\leq j\leq N)$. Table~\ref{chap3_tab-1} shows a single representative sequence of measurements received when the above model is used for simulation.
 	\begin{table}[h]
 		\caption{Received measurements with $\Theta=0.5$ and $N=2$}
 		\label{chap3_tab-1}
 		\begin{center}
 			\begin{tabular}{|p{0.2cm}|p{0.2cm}|c|c|p{1.05cm}|c|c|c|c|p{0.2cm}|c|}
 				\hline
 				$k$ & 1&2&3&4&5&6&7&8&9&10\\
 				\hline
 				$j$&$0$&$1$&$0$&$\beta_k^j=0$&$0$&$1$&$0$&$2$&$0$&$0$\\
 				\hline
 				$y_k$&$z_1$&$z_1$&$z_2$&lost$(y_3)$&$z_5$&$z_5$&$z_7$&$z_6$&$z_9$&$z_{10}$\\
 				\hline	
 			\end{tabular}
 		\end{center}
 	\end{table}
 	The main difference lies in the fact that this kind of models allow the measurements to be dropped when $\alpha_k^j$ are not $1$ for any value of $j$. For example, the models in \cite{sun2009linear, sun2010optimal, tiwari2020particle} have incorporated the scope for the packet drops along with the random delays. On the other hand, the authors in \cite{song2016linear, ma2011optimal} have used an additional Bernoulli variable in the structure of \cite{huang2015particle}  to model the missing instances in received measurements. However, in all the models, the same measurement can be received more than once, which is redundant and might be uninformative in reconstructing the states.
 \end{enumerate}
 The following remarks can be made in context of the existing delay models:    
 \begin{itemize}	
 	\item When the same measurements are received repetitively, they do not remain conditional independent of each other and the measurement noise sequence gets correlated over time. These phenomena restrict the estimator from using the standard Bayesian estimation structure.
 	\item The author in \cite{sun2012optimal} has proposed a model where measurements are not received more than once; however, they have used a set of Bernoulli random variables for different delay steps, which resulted in a visibly complicated and bulky expression.
 	\item All the models in literature indicate that a measurement getting no delay is more likely than it goes through a non-zero steps of delay and further, the probability decreases with the increase in number of delay steps. This might be representing the scenario close to the most of the realistic cases, however, it remains a special case of a general model where any number of delay steps can dominate in a set of received measurements. 
 \end{itemize}
 \subsection{Proposed Measurement Model}
To mathematically represent a measurement model that overcomes the difficulties mentioned above, we propose to employ the Poisson distribution as follows. If $y_k$ is the measurement received at time step $k$, then, 
\begin{equation}\label{delay_model}
	y_k=\sum_{j_k=0}^N \beta_k^{j_k} z_{k-j_k},
\end{equation}
where  $\beta_k^{j_k}$ is defined as
\begin{equation}\label{beta}
	\beta_k^{j_k}=
	\begin{cases}
		\alpha_k^0;  &j_k=0\\
		\alpha_k^{j_k} \prod_{i=1}^{j_k} (1-\alpha_{k-i}^{j_{k}-i}) ; & j_k>0.
	\end{cases}
\end{equation}
Here, $\alpha_k^{j_k} (j_k=0, 1, \cdots, N)$ is a set of binary variables with $N$ as the maximum value of its time-varying index, $j_k$; $j_k$ indicates the number of delay steps and if a measurement is received with a delay step greater than $N$, it is assumed to be uninformative for the estimation of states and considered as a packet drop at that step. Moreover, only one of $\alpha_k^{j_k}$, at most, takes value $1$ at time step $k$, and the rest of them are $0$. The Poisson distributed index, $j_k$, for which $\alpha_k^{j_k}$ holds a value of $1$, has the probability mass function (pmf) given as 
\begin{equation}\label{poiss_pmf}
	P(\alpha_k^{j_k}=1)=e^{-\lambda_k} \dfrac{\lambda^{j_k}_k}{j_k!},
\end{equation}   
where $\lambda_k$ is the mean of Poisson random variable $j_k$, i.e. $\mathbb{E}[j_k]=\lambda_k$, and represents the average value of delay at each time step. Also, the expectation of $\alpha_k^{j_k}$ is given as $\mathbb{E}[\alpha_k^{j_k}]=e^{-\lambda_k} \dfrac{\lambda^{j_k}_k}{j_k!}$.
\begin{rem}  To depict a situation where the same measurements can be received repetitively, a special case of \eqref{delay_model} with $\beta_k^{j_k}=1, \ \forall j_k$, can be given by
\begin{equation} \label{first_form}
	y_k=z_{k-j_k}; \quad j_k\leq k,
\end{equation} 
where $j_k$ is an i.i.d. Poisson random variable and it represents the number of delay steps that a measurement at time step $k-j_k$ suffers before it is received at time step $k$. However, this model does not restrict the maximum number of delay step.
 \end{rem}
\begin{rem} 
	To circumvent these characteristics of repetitive measurements and unrestricted number of delay steps in \eqref{first_form}, a set of binary variables, $\alpha_k^{j_k} (j_k=0, 1, \cdots, N)$ has been considered, where $N$ is the user-defined maximum number of permissible delay steps.\end{rem} 

\begin{figure}[h!]
	\centering
	\includegraphics[width=3.2in, height=2.1in]{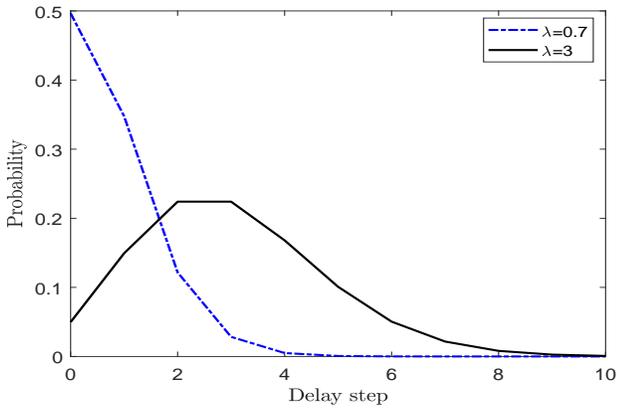}
	\caption{ Delay steps versus probability for Poisson distribution}
	\label{chap2_fig1}
\end{figure}

\begin{rem} 
 The proposed model in \eqref{delay_model} represents a generalized class of non-repetitive measurement models. Similarly, the expression in \eqref{first_form} with the introduction of the maximum number of permissible delays, $N$, depicts a generalized case of the repetitive measurement models with random packet drops.
\end{rem}
At any time step $k$, a measurement is delayed by $j_k$ steps with probability $P(\beta_k^{j_k}=1)$, and the probability of the same measurement getting dropped is $1-P(\sum_{j_k=0}^N\beta_k^{j_k}=1)$. These probabilities are calculated in Lemmas~\ref{prob_lemma} and \ref{prob_lemma2}.
\begin{lemma}\label{prob_lemma}
	The probability of a received measurement, $y_k$, being delayed by $j_k$ steps is 
	\begin{equation}\label{delay_prob}
		\gamma_k^{j_k} =\begin{cases}
			e^{-\lambda_k};&j_k=0\\
			e^{-\lambda_k}\dfrac{\lambda_k^{j_k}}{j_k!}\prod_{i=1}^{j_k}\left(1-e^{-\lambda_{k-i}}\dfrac{\lambda_{k-i}^{j_k-i}}{(j_k-i)!}\right);&j_k>0.\end{cases}
	\end{equation}
\end{lemma} 
\begin{proof}
	\textit{Case I ($j_k=0$):} By using Eqs.~\eqref{poiss_pmf} and \eqref{beta}, we have
	\begin{equation*}
		\begin{split}
			\gamma_k^{0}=P(\beta_k^{0}=1)&=\mathbb{E}[\beta_{k}^0=\alpha_k^{0}]\\
			&=e^{-\lambda_k}.
		\end{split}
	\end{equation*}
	\textit{Case II ($j_k>0$):} Similarly, $$\gamma_k^{j_k}=P(\beta_k^{j_k}=1)=\mathbb{E}\left[\alpha_k^{j_k} \prod_{i=1}^{j_k} (1-\alpha_{k-i}^{j_{k}-i})\right].$$ Considering the binary variables, $\alpha_k^{j_k} (j_k=0, 1, \cdots, N)$, are independent over the time steps, we can write 
	\begin{equation}
		\begin{split}
			\gamma_k^{j_k}&=\mathbb{E}[\alpha_k^{j_k}] \prod_{i=1}^{j_k} \mathbb{E}[(1-\alpha_{k-i}^{j_{k}-i})]\\
			&=\mathbb{E}[\alpha_k^{j_k}]\prod_{i=1}^{j_k}(1-\mathbb{E}[\alpha_{k-i}^{j_{k}-i}]).
		\end{split}
	\end{equation} 
	Using \eqref{poiss_pmf} in the above equation establishes \eqref{delay_prob}.
\end{proof}
\begin{lemma}\label{prob_lemma2}
	The probability that a measurement, $y_k$, is never received at the estimator side is given as $1-\sum_{j_k=0}^{{N}} \gamma_k^{j_k}$.	
\end{lemma}
\begin{proof}
	The probability that $\beta_k^{j_k}$ is zero for all permissible $j_k$ is given as
	\begin{equation*}
		\begin{split}
			P\left(\sum_{j_k=0}^N\beta_k^{j_k}=0\right)&=1-\sum_{j_k=0}^N\mathbb{E}[\beta_k^{j_k}]\\
			&=1-\sum_{j_k=0}^{{N}} \gamma_k^{j_k}.
		\end{split}
	\end{equation*}
\end{proof}
The proposed delay model has the following properties in comparison with the existing models:
\begin{enumerate}[(i)]
	\item Once we define a value for $N$, it models the random packet drops naturally with non-zero probability, which resembles a real scenario closely. A measurement is treated as lost when it goes through a delay more than $N$ steps.
	The delay models used in \cite{wang2013gaussian, huang2015particle, hermoso2009unscented} do not offer the scope for a measurement packet to get dropped and the estimator mandatorily receives a measurement at each time step. The work in \cite{ma2011optimal} has to use an additional random variable to represent the random measurement dropouts. 
	\item In the proposed model, one measurement is transmitted only once and no measurement is received more than once. Using the same measurement multiple times for the estimation is redundant and might not help in reconstructing the states at that time step. 
	The delay models in \cite{huang2015particle, wang2013gaussian, hermoso2009unscented} prefer the same measurement to be received more than once over the packet drop. Whereas in \cite{singh2016modified, tiwari2020particle}, the authors use the previously received measurement for estimating the state if there occurs a packet drop at the current time step.
	\item The loss of whiteness of the noise sequence prevents the designer from using the standard Bayesian estimation algorithms. However, in the proposed model, no measurement can be received more than once; as a result, the model maintains the conditional independence of the current measurement $y_k$ with respect to its previously received measurements $y_{1:k-1}$. Further, as shown in Appendix~\ref{uncorrelated_noise}, it keeps the whiteness of measurement noise sequences intact
	unlike the models used in \cite{huang2015particle, wang2013gaussian, hermoso2009unscented, singh2016modified, tiwari2020particle}. 
	\item The parameter, $\lambda_k$, is an important part of the proposed model and it captures a wide variety of delay scenarios for the networked systems with proper selection of its value. The property that a measurement getting no delay is more likely than it goes through a non-zero steps of delay is depicted with $\lambda_k \in (0,1]$. If we select $\lambda_k=3$, the measurements with delay steps equal to $3$ will be more likely and the probability decreases as we move along the either side of $3$. Fig. \ref{chap2_fig1} shows the probability versus number of delay steps for Poisson distributed $j_k$. Table~\ref{chap2_tab1} shows a batch of the received measurements when $\lambda_k=0.7$ for all values of $k$ with $N=2$.
\end{enumerate}
\begin{table}[h]
	\caption{Received measurements with $\lambda=0.7$ and $N=2$}
	\label{chap2_tab1}
	\begin{center}
		\begin{tabular}{|p{0.2cm}|p{0.2cm}|p{.7cm}|p{0.2cm}|p{0.2cm}|p{0.2cm}|p{.7cm}|p{0.2cm}|p{0.2cm}|p{0.2cm}|c|}
			\hline
			$k$ & 1&2&3&4&5&6&7&8&9&10\\
			\hline
			$j_k$ &$0$&$\beta_{2}^{j_2}=0$&$1$&$1$&$0$&$\beta_{6}^{j_6}=0$&$0$&$2$&$0$&$0$\\
			\hline
			$y_k$&$z_1$&lost ($\hat{y}_2$)&$z_2$&$z_3$&$z_5$&lost ($\hat{y}_{6}$)&$z_7$&$z_6$&$z_9$&$z_{10}$\\
			\hline	
		\end{tabular}
	\end{center}
\end{table}
\subsection{Problem Statement} We seek to design the Bayesian estimation algorithms employing the Gaussian approximation and sequential Monte Carlo (SMC) methods to recursively reconstruct the posterior density, $p(x_{k}|y_{1:k})$, and the expectation of a posterior density-integrable function  of the unobserved state, $\{x_k; \ k \in \mathbb{N}\}$, by using the received measurements, $\{y_k; \ k \in \mathbb{N}\}$. The measurements received are specified by the expression in \eqref{ideal_meas} and the proposed delay model in \eqref{delay_model}, and the unobserved states follow the dynamics given in \eqref{state_dynamics}.
\section{Gaussian Filters for Randomly Delayed Measurements}
In this section, we derive the nonlinear filtering algorithm under the Gaussian assumption for the proposed model \eqref{delay_model} by using the Bayesian framework. The joint density of states conditioned on the received measurements can be expressed as
\begin{equation}\label{joint_den}
	\begin{split}
		p(x_{0:k}|y_{1:k})&=\dfrac{p(x_{0:k},y_{1:k})}{p(y_{1:k})}\\
		&=\dfrac{p(y_k|x_{0:k},y_{1:k-1})p(x_{0:k},y_{1:k-1})}{p(y_k|y_{1:k-1})p(y_{1:k-1})}\\
		&=\dfrac{p(y_k|x_{0:k},y_{1:k-1})p(x_{0:k}|y_{1:k-1})}{p(y_k|y_{1:k-1})}.
	\end{split}
\end{equation}
From the measurement models \eqref{ideal_meas} and \eqref{delay_model}, we can see that the received measurement, $y_k$, is correlated with the states, $x_k, \cdots, x_{k-N}$. Hence, relaxing the standard assumption of independent measurement, we can consider that $y_k$, conditioned on $x_{k-N:k}$, is independent of the previous measurements and states, i.e. $p(y_k|x_{0:k},y_{1:k-1})=p(y_k|x_{k-N:k})$. Also, for a recursive estimation at time step $k$, it is assumed that the estimate of states up to time step $k-1$ is already known. Thus, the filtering density of state, by using \eqref{joint_den}, can be given as
\begin{equation}\label{filtering_den}
	p(x_k|y_{1:k})=\dfrac{p(y_k|x_{k-N:k})p(x_k|y_{1:k-1})}{p(y_k|y_{1:k-1})},
\end{equation} 
where the predictive density, $p(x_k|y_{1:k-1})$, is given by the Chapman-Kolmogorov integral as
\begin{equation}\label{predictive_den}
	p(x_k|y_{1:k-1})=\int p(x_k|x_{k-1})p(x_{k-1}|y_{1:k-1})\mathrm{d}x_{k-1}.
\end{equation}
Let us Consider that the process noise, $\eta_{k-1}$, and the measurement noise, $v_k$, are a zero mean, white Gaussian sequence with covariances, $Q_{k-1}$ and $R_k$, respectively. The initial state, $x_0$, also follows the Gaussian distribution, and $x_0$, $\eta_{k-1}$ and $v_k$ are uncorrelated sequences. Now, assuming that the predictive density, $p(x_k|y_{1:k-1})$, in \eqref{predictive_den}  is Gaussian if $p(x_{k-1}|y_{1:k-1})$ is Gaussian, the first and second moments of $p(x_k|y_{1:k-1})$ can be given as
\begin{equation}\label{prediction}
	\begin{split}
	\hat{x}_{k|k-1}&=\mathbb{E}[x_k|y_{1:k-1}] \\ \mathbf{P}_{k|k-1}&=\mathbb{E}[(x_k-\hat{x}_{k|k-1})(x_k-\hat{x}_{k|k-1})^\top|y_{1:k-1}].
	\end{split}
\end{equation}
Since the prediction density in \eqref{predictive_den} depends on the state dynamics and the previous estimate and not on the current measurement, the expectation over it can be computed with any Gaussian approximation methods available in literature \cite{haykin2009cubature,julier1997new,ito}.

Similarly, assume that the predictive density of the current measurement is also Gaussian i.e., $p(y_k|y_{1:k-1})=\mathcal{N}(y_k; \hat{y}_{k|k-1}, \mathbf{P}_{k|k-1}^{yy})$. Its moments are given in Lemma~\ref{expec_y}, where $\hat{z}_{k-j_k|k-1}$ and $\mathbf{P}_{k-j_k|k-1}^{zz}$ can be computed as illustrated in the Gaussian approximation methods.
\begin{lemma}\label{expec_y}
	The predicted estimate of measurement at time step $k$ is 
	\begin{equation}\label{exp_y}
		\hat{y}_{k|k-1}=\sum_{j_k=0}^{N} \bar{\gamma}_k^{j_k} \hat{z}_{k-j_k|k-1},
	\end{equation}
and the measurement covariance is given by
\begin{equation}\label{meas_cov}
	\begin{split}
	\mathbf{P}_{k|k-1}^{yy}&=\sum_{j_k=0}^N \bar{\gamma}_k^{j_k}\mathbf{P}_{k-j_k|k-1}^{zz}\\&\quad + \sum_{j_k=0}^N\bar{\gamma}_k^{j_k}(1-\bar{\gamma}_k^{j_k})\hat{z}_{k-j_k|k-1}\hat{z}_{k-j_k|k-1}^\top .
	\end{split}
\end{equation}
\end{lemma}
\begin{proof}
The predicted estimate of received measurement at $k$th time step can be given as
\begin{equation*}
\hat{y}_{k|k-1}=\mathbb{E}[y_k|y_{1:k-1}]=\mathbb{E}\left[\sum_{j_k=0}^N\beta_k^{j_k} z_{k-j_k}|y_{1:k-1}\right].
\end{equation*} 
Since the variables $\beta_k^{j_k}$ and $z_{k-j_k}$ are independent, we can write above expectation as
\begin{equation*}
	\hat{y}_{k|k-1}=\sum_{j_k=0}^N\mathbb{E}[\beta_k^{j_k}|y_{1:k-1}] \mathbb{E}[z_{k-j_k}|y_{1:k-1}].
\end{equation*} 
Using the fact that $\beta_k^{j_k}$ and the past received measurements, $y_{1:k-1}$, are uncorrelated, and by Lemma~\ref{prob_lemma}, the above equation leads to \eqref{exp_y}. Note that this estimate of measurement is computed excluding the time instants when no measurement is received and hence, the normalized delay probability, $\bar{\gamma}_k^{j_k}=\gamma_k^{j_k}/\sum_{j_k=0}^N\gamma_k^{j_k}$, is used as the expectation of $\beta_k^{j_k}$. 

Further, from \eqref{exp_y}, the estimated error in received measurement at $k$th time step can be given as
\begin{equation}\label{m1m2}
	\begin{split}
	y_k-\hat{y}_{k|k-1}&=\sum_{j_k=0}^N\beta_k^{j_k} z_{k-j_k}-\sum_{j_k=0}^{N} \bar{\gamma}_k^{j_k} \hat{z}_{k-j_k|k-1}\\
	&=M_1-M_2,
	\end{split}
\end{equation}
and the measurement covariance, using \eqref{m1m2}, is defined as
\begin{equation}\label{expects}
	\begin{split}
	\mathbf{P}_{k|k-1}^{yy}&=\mathbb{E}[(y_k-\hat{y}_{k|k-1})(y_k-\hat{y}_{k|k-1})^\top|y_{1:k-1}]\\
	&=\mathbb{E}[M_1M_1^\top|y_{1:k-1}]-\mathbb{E}[M_1M_2^\top|y_{1:k-1}]\\&\quad-\mathbb{E}[M_2M_1^\top|y_{1:k-1}]+\mathbb{E}[M_2M_2^\top|y_{1:k-1}].
	\end{split}
\end{equation}
Now, we compute the expectations of \eqref{expects} as follows:
\begin{equation*}
	\begin{split}
		&\mathbb{E}[M_1M_1^\top|y_{1:k-1}]\\&=\mathbb{E}\Big[\sum_{s=0}^N\beta_k^s z_{k-s}\sum_{l=0}^N\beta_k^l z_{k-l}^\top|y_{1:k-1}\Big]\\
		&=\sum_{s=0}^N\sum_{l=0}^N \mathbb{E}[\alpha_k^s(1-\alpha_{k-1}^{s-1})\cdots (1-\alpha_{k-s}^0)\alpha_k^l(1-\alpha_{k-1}^{l-1})\cdots\\
		&\quad\times (1-\alpha_{k-l}^0)]\mathbb{E}[z_{k-s}z_{k-l}^\top]
	\end{split}
\end{equation*}
	Consider the following cases for the computation of above equation.\\
	\underline{\emph{Case-I ($s=l$)}}: Given that $\alpha_{k-s}^s$ is a binary variable and $\mathbb{E}[(\alpha_{k-s}^s)^2]=\mathbb{E}[\alpha_{k-s}^s]$, and using the relation that $\mathbf{P}_{k-s|k-1}^{zz}=\mathbb{E}[z_{k-s}z_{k-s}^\top]-\hat{z}_{k-s|k-1}\hat{z}_{k-s|k-1}^\top$, we have
	\begin{equation}\label{1a}
		\begin{split}
				&\mathbb{E}[M_1M_1^\top|y_{1:k-1}]\\&= \sum_{s=0}^N] \mathbb{E}[(\alpha_k^s)^2(1-\alpha_{k-1}^{s-1})^2\cdots (1-\alpha_{k-s}^0)^2\mathbb{E}[z_{k-s}z_{k-s}^\top]\\
				&=\sum_{s=0}^N\mathbb{E}[\beta_k^s](\mathbf{P}_{k-s|k-1}^{zz}+\hat{z}_{k-s|k-1}\hat{z}_{k-s|k-1}^\top)\\
				&=\sum_{s=0}^N\bar{\gamma}_k^s(\mathbf{P}_{k-s|k-1}^{zz}+\hat{z}_{k-s|k-1}\hat{z}_{k-s|k-1}^\top).
		\end{split}
	\end{equation}
\underline{\emph{Case-II ($s\neq l$)}}: Using the fact that the non-delayed measurements, $z_k (k\in \mathbb{N})$, and $\alpha_k^j (k\in \mathbb{N}, 0\leq j\leq N)$ are independent, we can write
\begin{equation}\label{1b}
	\begin{split}
		&\mathbb{E}[M_1M_1^\top|y_{1:k-1}]\\&=\sum_{s=0}^N\sum_{l=0}^N \mathbb{E}[\alpha_k^s(1-\alpha_{k-1}^{s-1})\cdots (1-\alpha_{k-s}^0)]\mathbb{E}[\alpha_k^l(1-\alpha_{k-1}^{l-1})\\
		&\quad\times\cdots (1-\alpha_{k-l}^0)]\mathbb{E}[z_{k-s}]\mathbb{E}[z_{k-l}^\top]\\
		&=\sum_{s=0}^N\sum_{l=0}^N \bar{\gamma}_k^s\bar{\gamma}_k^l \hat{z}_{k-s|k-1}\hat{z}_{k-l|k-1}^\top.
	\end{split}
\end{equation}
Again, computing the second term of \eqref{expects}, we can write
\begin{equation}\label{2}
	\begin{split}
		&\mathbb{E}[M_1M_2^\top|y_{1:k-1}]\\&=\mathbb{E}\Big[\sum_{s=0}^N\beta_k^sz_{k-s}\sum_{l=0}^N\bar{\gamma}_k^l\hat{z}_{k-l|k-1}^\top|y_{1:k-1}\Big]\\
		&=\mathbb{E}\Big[\sum_{s=0}^N\beta_k^sz_{k-s}|y_{1:k-1}\Big]\sum_{l=0}^N\bar{\gamma}_k^l\hat{z}_{k-l|k-1}^\top|y_{1:k-1}\\
		&=\sum_{s=0}^N\sum_{l=0}^N\bar{\gamma}_k^s\bar{\gamma}_k^l\hat{z}_{k-s|k-1}\hat{z}_{k-l|k-1}^\top.
	\end{split}
\end{equation}
Similarly,
\begin{equation}\label{3}
	\mathbb{E}[M_2M_1^\top|y_{1:k-1}]=\sum_{s=0}^N\sum_{l=0}^N\bar{\gamma}_k^s\bar{\gamma}_k^l\hat{z}_{k-s|k-1}\hat{z}_{k-l|k-1}^\top.
\end{equation}
Lastly, \begin{equation}\label{4}
	\mathbb{E}[M_2M_2^\top|y_{1:k-1}]=\sum_{s=0}^N\sum_{l=0}^N\bar{\gamma}_k^s\bar{\gamma}_k^l\hat{z}_{k-s|k-1}\hat{z}_{k-l|k-1}^\top.
\end{equation}
Now, if we substitute the equations \eqref{1a}, \eqref{1b}, \eqref{2}, \eqref{3}, and \eqref{4} into \eqref{expects}, it establishes \eqref{meas_cov}. 
\end{proof}
 Proceeding further to obtain the posterior estimate, the cross-covariance, $\mathbf{P}_{k|k-1}^{xy}$, is derived in the following lemma.
\begin{lemma}
	The cross-covariance at time step $k$ is given as
	\begin{equation}\label{cross-cov}
		\mathbf{P}_{k|k-1}^{xy}=\sum_{s=0}^N \bar{\gamma}_k^s \mathbf{P}_{k,k-s|k-1}^{xz}.
	\end{equation}
\end{lemma}
\begin{proof}
	Using \eqref{delay_model} and \eqref{exp_y}, the conditional cross-covariance is defined as
	\begin{equation*}
		\begin{split}
			\mathbf{P}_{k|k-1}^{xy}&=\mathbb{E}[(x_k-\hat{x}_{k|k-1})(y_k-\hat{y}_{k|k-1})|y_{1:k-1}]\\
			&=\mathbb{E}\Big[(x_k-\hat{x}_{k|k-1})\Big(\sum_{s=0}^N \beta_k^s z_{k-s}-\sum_{s=0}\bar{\gamma}_k^s\hat{z}_{k-s|k-1}\Big)\Big]\\
			&=\mathbb{E}\Big[(x_k-\hat{x}_{k|k-1})\Big(\sum_{s=0}^N\beta_k^s(z_{k-s}-\hat{z}_{k-s|k-1})\\&\quad+\sum_{s=0}^N \beta_k^s\hat{z}_{k-s|k-1}-\bar{\gamma}_k^s\hat{z}_{k-s|k-1}\Big)\Big]\\
			&=\sum_{s=0}^N\mathbb{E}[\beta_k^s]\mathbb{E}[(x_k-\hat{x}_{k|k-1})(z_{k-s}-\hat{z}_{k-s|k-1})]\\&\quad
			+\sum_{s=0}^N\mathbb{E}[\beta_k^s-\bar{\gamma}_k^s]\mathbb{E}[(x_k-\hat{x}_{k|k-1})\hat{z}_{k-s|k-1}].
		\end{split}
	\end{equation*}
Since $\mathbb{E}[\beta_k^s-\bar{\gamma}_k^s]=0$ and $\mathbb{E}[(x_k-\hat{x}_{k|k-1})(z_{k-s}-\hat{z}_{k-s|k-1})]=\mathbf{P}_{k,k-s|k-1}^{xz}$, we have
\begin{equation*}
	\mathbf{P}_{k|k-1}^{xy}=\sum_{s=0}^N \bar{\gamma}_k^s \mathbf{P}_{k,k-s|k-1}^{xz}.
\end{equation*}
\end{proof}	
	\begin{rem}
		Under the Gaussian assumption, $\mathbf{P}_{k,k-s|k-1}^{xz}$ in \eqref{cross-cov} is given by
		\begin{equation*}
			\begin{split}
			\mathbf{P}_{k,k-s|k-1}^{xz}=&\int x_kh_{k-s}(x_{k-s})\mathcal{N}(x_k; \hat{x}_{k|k-1}, \mathbf{P}_{k|k-1})\\&\quad\times\mathrm{d}x_k-\hat{x}_{k|k-1}\hat{z}_{k-s|k-1},
			\end{split}
		\end{equation*}
	where the above integration can be approximated by a Gaussian approximation method available in literature.
	\end{rem}
\begin{theorem}
	The posterior estimate and covariance for the system \eqref{state_dynamics},\eqref{ideal_meas},\eqref{delay_model} are given as
	\begin{align}
		\hat{x}_{k|k}&=\hat{x}_{k|k-1} + K_k(y_k-\hat{y}_{k|k-1}), \label{post_mean}\\ 
		\mathbf{P}_{k|k}&=\mathbf{P}_{k|k-1}-K_k\mathbf{P}_{k|k-1}^{yy}K_k^\top,	\label{post_cov}	
	\end{align}
	where $K_k=\mathbf{P}_{k|k-1}^{xy}(\mathbf{P}_{k|k-1}^{yy})^{-1}.$
\end{theorem}
\begin{proof}
	Proceeding to computing the posterior filtering density, Eq.~\eqref{filtering_den} can be rewritten as
	\begin{equation}\label{bayes_joint}
		p(x_k|y_{1:k})=\dfrac{p(y_k,x_k|y_{1:k-1})}{p(y_k|y_{1:k-1})},
	\end{equation}
	where the joint density, $p(y_k,x_k|y_{1:k-1})$, is Gaussian under our earlier assumption about the predictive densities and can be given as
	\begin{equation}\label{joint_gauss}
		\begin{split}
			&p(y_k,x_k|y_{1:k-1})
			\\&=\mathcal{N}\begin{pmatrix} \begin{bmatrix} x_k\\y_k \end{bmatrix};& \begin{bmatrix} \hat{x}_{k|k-1}\\ \hat{y}_{k|k-1} \end{bmatrix},
				&\begin{bmatrix} \mathbf{P}_{k|k-1} & \mathbf{P}_{k|k-1}^{xy}\\ (\mathbf{P}_{k|k-1}^{xy})^\top & \mathbf{P}_{k|k-1}^{yy} \end{bmatrix} \end{pmatrix},
		\end{split}
	\end{equation}
	where the covariances, $\mathbf{P}_{k|k-1}$, $\mathbf{P}_{k|k-1}^{yy}$, and $\mathbf{P}_{k|k-1}^{xy}$ are defined in \eqref{prediction},\eqref{meas_cov}, and \eqref{cross-cov} respectively.
	Now, substituting \eqref{joint_gauss} into \eqref{bayes_joint} and performing the squaring operation for a Gaussian density (see Appendix A of \cite{wang2013gaussian}), we have
	\begin{equation}
		\begin{split}
			p(x_k|y_{1:k})=\mathcal{N}(x_k; \hat{x}_{k|k}, \mathbf{P}_{k|k}),
		\end{split}
	\end{equation}
	where $\hat{x}_{k|k}$ and $\mathbf{P}_{k|k}$ are given in \eqref{post_mean} and \eqref{post_cov}, respectively.
\end{proof}
Thus, \eqref{prediction} and \eqref{post_mean} present the predicted and posterior estimates, respectively, for a stochastic system under the Gaussian assumption. Further, if the measurement is not received at a time step $k$, we use the predicted measurement, $\hat{y}_{k|k-1}$, for the state estimation.
\section{ SMC Method for Randomly Delayed Measurements} 
In this section, we develop an estimation algorithm without assuming a particular distribution for the system noises and prior information. Consider $\{x_k; \ k \in \mathbb{N}\}$ be an unobserved Markov process with an initial distribution $p(x_o)$ and the transitional density specified by \eqref{state_dynamics}. The received measurements, $\{y_k; \ k \in \mathbb{N}\}$, are conditionally independent given the process $\{x_k; \ k \in \mathbb{N}\}$ with the likelihood density defined by  \eqref{ideal_meas} and \eqref{delay_model}. The posterior distribution $p(x_{0:k}|y_{1:k})$ can be approximated with the help of a set of i.i.d. samples drawn from the distribution as \cite{doucet2009tutorial}
\begin{equation}\label{posterior_particle}
	\hat{p}(x_{0:k}|y_{1:k})=\dfrac{1}{N_s}\sum_{i=1}^{N_s}\delta_{x_{0:k}^i}(x_{0:k}),
\end{equation}
where $N_s$ is the total number of samples and the particles, $\{x_{0:k}^i\}_{i=1}^{N_s}$ are drawn from the posterior distribution. Unfortunately, the posteriors are usually non-standard, multivariate and known only up to a proportional constant and hence the sampling of particles is almost impossible. Alternatively, we adopt a Bayesian importance sampling method, where we select a known and easy-to-sample proposal distribution, $q(x_{0:k}|y_{1:k})$, from which the particles can easily be drawn. If $g_k(x_{0:k})$ is a $p(x_{0:k}|y_{1:k})$-integrable function, the expectation, $\mathbb{E}_{p(\cdot|y_{1:k})}(g_k(x_{0:k}))$, can be given as
\begin{equation}\label{expect}
\begin{split}
&\mathbb{E}_{p(\cdot|y_{1:k})}(g_k(x_{0:k}))\\&=\int g_k(x_{0:k})\dfrac{ p(x_{0:k}|y_{1:k})}{q(x_{0:k}|y_{1:k})}q(x_{0:k}|y_{1:k})dx_{0:k}\\	
&=\int g_k(x_{0:k})\dfrac{ p(y_{1:k}|x_{0:k})p(x_{0:k})}{p(y_{1:k})q(x_{0:k}|y_{1:k})}q(x_{0:k}|y_{1:k})dx_{0:k}\\
&=\dfrac{1}{p(y_{1:k})} \int g_k(x_{0:k})w_k(x_{0:k})q(x_{0:k}|y_{1:k})dx_{0:k},
\end{split}
\end{equation}
where $w_k(x_{0:k})$ is the unnormalized importance weight defined as
\begin{equation}\label{imp_wt}
	w_k=\dfrac{p(y_{1:k}|x_{0:k})p(x_{0:k})}{q(x_{0:k}|y_{1:k})}.
\end{equation} 
Now, to compute $\mathbb{E}_{p(\cdot|y_{1:k})}(g_k(x_{0:k}))$ in terms of the expectations that are taken over the proposal distribution, $q(x_{0:k}|y_{1:k})$, we can write the normalizing constant as $p(y_{1:k})=\int p(y_{1:k}|x_{0:k})p(x_{0:k})dx_{0:k}$ and substitute it in  \eqref{expect}. After some rearrangements, Eq. \eqref{expect} can be written as
\begin{equation*}
\mathbb{E}_{p(\cdot|y_{1:k})}(g_k(x_{0:k}))=\dfrac{\mathbb{E}_{q(\cdot|y_{1:k})}(w_k(x_{0:k})g_k(x_{0:k}))}{\mathbb{E}_{q(\cdot|y_{1:k})}(w_k(x_{0:k}))}.
\end{equation*}
The estimate of the above expectations can be computed with the help of a set of i.i.d. samples, $\{x_{0:k}^i\}_{i=1}^{N_s}$ drawn from the proposal distribution, $q(x_{0:k}|y_{1:k})$, and Eq. \eqref{imp_wt}. Further, our aim is to sequentially estimate the posterior distribution and the associated expectations at each time step $k$. In order to achieve the sequential estimation, the proposal distribution can be assumed to be decomposed as
\begin{equation}\label{proposal_fact}
	q(x_{0:k}|y_{1:k})=q(x_{0:k-1}|y_{1:k-1})q(x_k|x_{0:k-1},y_{1:k}).
\end{equation}
Here, we have used the chain rule and assumed that the states ($x_{0:k-1}$) are independent of the future measurement ($y_k$). Also, from the measurement models \eqref{ideal_meas} and \eqref{delay_model}, it is evident that the current measurement, $y_k$, is correlated with the states, $x_k, x_{k-1}, \cdots,$ and $ x_{k-N}$. Hence, by using chain rule and under our assumptions that the states are a Markov process and the measurements, conditioned on the states, are  independent, we have
\begin{equation}\label{factor}
	\begin{split}
		p(x_{0:k})&=p(x_0)\prod_{l=1}^k p(x_l|x_{l-1}),\\
		p(y_{1:k}|x_{0:k})&=\prod_{l=1}^k p(y_l|x_{l-\bar{N}:l});\ k>0,
	\end{split}
\end{equation}
where $\bar{N}=\text{min}(N,l-1)$. Now, substituting Eqs. \eqref{proposal_fact} and \eqref{factor} into Eq. \eqref{imp_wt}, a recursive expression for the unnormalized importance weight can be derived as
\begin{equation}\label{importance_weight}
	\begin{split}
			w_k&=\dfrac{p(y_{1:k}|x_{0:k})p(x_{0:k})}{q(x_{0:k-1}|y_{1:k-1})q(x_k|x_{0:k-1},y_{1:k})}\\
			&=w_{k-1}\dfrac{p(y_{1:k}|x_{0:k})p(x_{0:k})}{p(y_{1:k-1}|x_{0:k-1})p(x_{0:k-1})q(x_k|x_{0:k-1},y_{1:k})}\\
			&=w_{k-1}\dfrac{p(y_{k}|x_{k-\bar{N}:k})p(x_k|x_{k-1})}{q(x_k|x_{0:k-1},y_{1:k})}.
	\end{split}
\end{equation}
\subsection{Estimation of State Posterior Density}
Now, the whole set of particles is divided into as many groups as the different delay steps for the received measurement. Each group represents a probable version of measurement with a certain number of delay steps. Thus, instead of one set of particles, we can then use those groups of particles to approximate the state posterior pdf.
\begin{theorem}\label{chap3_post_approx}
	The filtering density, $p(x_k|y_{1:k})$, for the system \eqref{state_dynamics}, \eqref{ideal_meas}, \eqref{delay_model} can be computed with the help of a set of i.i.d. samples drawn from the proposal density, $q(x_k|x_{0:k-1}^i,y_{1:k})$, as
	\begin{equation}\label{particle_approx}
		\hat{p}(x_k|y_{1:k})=\sum_{j_k=0}^{\bar{N}}\sum_{i=1}^{N_s^{j_k}}\bar{w}_k^{j_k,i}\delta_{x_{k}^{j_k,i}}(x_{k}),
	\end{equation}
	where $w_k^{j_k,i}=w_{k-1}^{j_{k-1},i}p(y_{k}|x_{k-j_k}^i)\dfrac{p(x_k^i|x_{k-1}^i)}{q(x_k^i|x_{0:k-1}^i,y_{1:k})}$, \ $\bar{w}_k^{j_k,i}=\dfrac{{w}_k^{j_k,i}}{\sum_{j_k=0}^{\bar{N}}\sum_{i=1}^{N_s^{j_k}}{w}_k^{j,i}}$,  and \ $N_s^{j_k}=\bar{\gamma}_k^{j_k} N_s$ such that \ $\sum_{j_k=0}^{\bar{N}}N_s^{j_k} = N_s$.
\end{theorem}
\begin{proof}
	In Eq.~\eqref{importance_weight},  the likelihood density, $p(y_k|x_{k-\bar{N}:k})$, can be written as the joint density, $p(y_k,\beta_k^{j_k}|x_{k-\bar{N}:k})$, which is marginalized over all the possible values of $\beta_k^{j_k}\ (j_k=0, \cdots, \bar{N})$, as follows.
	\begin{equation}\label{joint_p}
		\begin{split}
			p(y_k|x_{k-\bar{N}:k})=\sum_{j_k=0}^{\bar{N}} p(y_k,\beta_k^{j_k}|x_{k-\bar{N}:k})
		\end{split}
	\end{equation}
	At most, one of $\beta_k^{j_k} \ (j_k=0,\cdots, \bar{N})$ is $1$ at any given time step $k$ and the others are zero with the probability given in \eqref{delay_prob}. Note that the likelihood is computed for the time steps we receive a measurement and hence, the combination where all $\beta_k^{j_k}$ are zero, which results into a measurement loss, is not considered. From \eqref{delay_model} and \eqref{delay_prob}, Eq.~\eqref{joint_p} can be expanded as follows:
	\begin{equation}\label{likelihood_fact}
		\begin{split}
			p(y_k|x_{k-\bar{N}:k})&=\sum_{j_k=0}^{\bar{N}}p(y_k|\beta_k^{j_k},x_{k-\bar{N}:k})p(\beta_k^{j_k}|x_{k-\bar{N}:k})\\
			&=p(y_k|\beta_k^0=1,x_{k-\bar{N}:k})P(\beta_k^0=1)+\cdots\\&\quad + p(y_k|\beta_k^{\bar{N}}=1,x_{k-\bar{N}:k})P(\beta_k^{\bar{N}}=1)\\
			&=\sum_{j_k=0}^{\bar{N}}p(y_{k}|x_{k-j_k})\bar{\gamma}_k^{j_k}.
		\end{split}
	\end{equation}
	Substituting \eqref{likelihood_fact} in \eqref{importance_weight}, we have
	\begin{equation}
		\begin{split}
			w_k&=w_{k-1}\sum_{j_k=0}^{\bar{N}}p(y_{k}|x_{k-j_k})\bar{\gamma}_k^{j_k}\dfrac{p(x_k|x_{k-1})}{q(x_k|x_{0:k-1},y_{1:k})}\\
			&=\sum_{j_k=0}^{\bar{N}}w_{k}^{j_k}\bar{\gamma}_k^{j_k},
		\end{split}
	\end{equation}
	where the recursive unnormalized importance weight, when the measurement is supposed to be delayed by $j_k$ steps, is
	$$w_k^{j_k}=w_{k-1}^{j_{k-1}}p(y_{k}|x_{k-j_k})\dfrac{p(x_k|x_{k-1})}{q(x_k|x_{0:k-1},y_{1:k})}.$$
	Now, using the i.i.d. samples to approximate the posterior distribution similar to \eqref{posterior_particle} when the particles are sampled from the proposal distribution, $q(x_k|x_{0:k-1}^i,y_{1:k})$, we have
	\begin{equation}
		\begin{split}
			\hat{p}(x_k|y_{1:k})&=\sum_{j_k=0}^{\bar{N}}\bar{\gamma}_k^j\sum_{i=1}^{N_s}\bar{w}_k^{j_k,i}\delta_{x_{k}^i}(x_{k})\\
			&=\sum_{j_k=0}^{\bar{N}}\sum_{i=1}^{N_s^{j_k}}\bar{w}_k^{j_k,i}\delta_{x_{k}^{j_k,i}}(x_{k}),
		\end{split}
	\end{equation}
	where $\bar{w}_k^{j_k,i}=\frac{{w}_k^{j_k,i}}{\sum_{j_k=0}^{\hat{N}}\sum_{i=1}^{N_s^{j_k}}{w}_k^{j_k,i}}$, and $N_s^{j_k}=\bar{\gamma}_k^{j_k} N_s$. Also, since $\sum_{j_k=0}^{\bar{N}}\bar{\gamma}_k^j = 1$, it implies that $\sum_{j_k=0}^{\bar{N}}N_s^{j_k} = N_s$.
\end{proof}
\begin{rem}
The computation of importance weight in Theorem~\ref{chap3_post_approx} is valid for the time instant when the system \eqref{state_dynamics}, \eqref{ideal_meas}, \eqref{delay_model} receives a measurement. However, when $\beta_k^{j_k}=0$ for all values of $j_k$ and the measurement is lost; there are only the particles sampled from the proposal density and $w_k^{j_k,i}=w_{k-1}^{j_{k-1},i}$. 
\end{rem}
\subsection{Delay transition rule for particles}
It is clear from Theorem~\ref{chap3_post_approx} that the particles are divided into $\bar{N}+1$ groups at any time step $k$, where each group supports the fact that the received measurement, $y_k$, is delayed through $j_k$ ($0\leq j_k \leq \bar{N}$) steps  and has a strength of $N_s^{j_k}$ particles. This necessitates framing of a set of rules for delay assignment to each particle after it has been drawn from the proposal distribution, $q(x_k|x_{0:k-1}^i,y_{1:k})$. Now, assume that the delay assignment to each particle upto time step $k-1$ is known, then, at step $k$, we need to consider two things to assign the delay step:
\begin{itemize}
	\item[(i)] the delay history of the particle for last $\bar{N}$ steps, and
	\item[(ii)]  the delay probability, $\bar{\gamma}_k^{j_k}$, for $0\leq j_k\leq\bar{N}$.
\end{itemize} 
Since a measurement can not be received more than once, the particle, $x_{k-\tau}^{j_k,i}$, which was assigned a delay of $j_{k-\tau}$ steps to support the measurement $z_{k-j_{k-\tau}}$ at time step $k-\tau$, can not support a measurement $z_{k-j_k}$ bearing a delay of $j_k$  at time step $k$ if $j_k=j_{k-\tau}+\tau, \ 0< \tau \leq \bar{N}$. This implies that the probability of $i$th particle at time step $k$ being assigned a delay of $j_k$ steps is given as
\begin{equation}\label{particle_delay}
	P(\beta_k^{j_k}=1|x_k^{j_k,i})=	\gamma_k^{j_k,i}=\begin{cases}
		0;\ \text{if}\  j_k=j_{k-\tau}+\tau, \\
		\bar{\gamma}_k^{j_k}; \ \text{else},\ \text{when}\ 0< \tau \leq \bar{N}.
	\end{cases}
\end{equation}
Further, the delay assignment for each particle can be executed as follows. The $i$th particle at time step $k$ is assigned a delay of $\tau$ steps if
$$ \sum_{d=0}^\tau \bar{\gamma}_k^{d,i} \leq u <  \sum_{d=0}^{\tau+1} \bar{\gamma}_k^{d,i}; \ 0\leq \tau \leq \bar{N},$$
where  $\bar{\gamma}_k^{d,i}=\dfrac{\gamma_k^{d,i}}{\sum_{d=0}^{\bar{N}}\gamma_k^{d,i}}$, and $u$ is a uniformly sampled number in $[0,1]$.
\begin{rem}
The sum of product of the likelihood densities method adopted in \cite{tiwari2020particle} and \cite{huang2015particle} includes every particle repeatedly for $\bar{N}$ steps to compute the importance weight irrespective of the fact that it has been already used to approximate the posterior state density at previous steps. On the other hand, this chapter excludes a particle from being used in the computation of importance weight if it once has been used in earlier steps as given in \eqref{particle_delay}. Hence, we present a method where the relevant particles get a higher chance to represent the posterior density. 
\end{rem}  
\subsection{Resampling}
Once the delay is assigned to every particle for the current step, the associated importance weight is computed as stated in Theorem~\ref{chap3_post_approx}. On the basis of computed importance weights, the particles are discretely resampled to select only those particles which support the current measurement with significant weights. The value of delay steps of the resampled particles actually carry the delay information of the received measurement. Heuristically, the probability of the measurement being delayed by $j_k$ steps can be approximately given as $\frac{N_s^{'j_k}}{N_s}$, where $N_s^{'j_k}$ is the number of particles assigned with $j_k$ steps delay after the resampling. Note that Lemma~\ref{prob_lemma} dictates about the prior probability of a measurement being delayed by certain steps, on the other hand, it is the posterior probability of delay that we calculate with the help of resampled state particles. A more systematic way of computing this posterior probability is illustrated in the following subsection. 
\subsection{Estimation of delay steps}
The delay variable $j_k$ is a Poisson i.i.d. random number, which is correlated with the states and received measurement through $\beta_k^{j_k}$ as given in \eqref{delay_model} and \eqref{ideal_meas}.
\begin{theorem}
	The filtering estimate of random delay, $\hat{j}_k$, for the system \eqref{state_dynamics}, \eqref{ideal_meas}, \eqref{delay_model} is given as\begin{equation}
		\hat{j}_k=\arg \max_{0\leq j_k \leq \bar{N}} \hat{p}(j_k|y_{1:k}),
	\end{equation}
	where $\hat{p}(j_k|y_{1:k})=\sum_{i=1}^{N_s^{j_k}} \bar{w}_k^{j_k,i} \bar{\gamma}_k^{j_k,i}$.
\end{theorem}
\begin{proof}
	The fact that $\beta_k^j$ and $y_{1:k-1}$ are uncorrelated, the predictive density of delay step, ${p}(j_k|y_{1:k-1})$, is $P(\beta_k^{j_k}=1)$. Then, the filter density can be given as
	\begin{equation*}
		\begin{split}
			{p}(j_k|y_{1:k})&=\int {p}(j_k,x_k|y_{1:k})\mathrm{d}x_k\\
			&=\int {p}(j_k|x_k)p(x_k|y_{1:k})\mathrm{d}x_k\\
			&=\int {P}(\beta_k^{j_k}=1|x_k)p(x_k|y_{1:k})\mathrm{d}x_k.
		\end{split}
	\end{equation*} 
	Now, if we use the particle approximation from \eqref{particle_approx} with $\beta_k^{j_k}=1$ i.e., $y_k=z_{k-j_k}$, then only the particles that have been assigned with a delay of $j_k$ steps is used to approximate the above integral. Using the probability of $i$th particle being assigned a delay of $j_k$ steps, we can further write the above equation as
	\begin{equation*}
		\begin{split}
			\hat{p}(j_k|y_{1:k})&=\sum_{i=1}^{N_s^{j_k}}\bar{w}_k^{j_k,i}p(\beta_k^{j_k}=1|x_k^{j_k,i})\\
			&=\sum_{i=1}^{N_s^{j_k}}\bar{w}_k^{j_k,i}\bar{\gamma}_k^{j_k,i}.
		\end{split}
	\end{equation*} 
	If ${p}(j_k|y_{1:k})$ is maximized over $0\leq j_k\leq \bar{N}$, we obtain our estimate for the delay step.
\end{proof}
\begin{corollary}
	If $d$ denotes the delay assigned to the $i$th particle at time step $k$, the mean of delay steps is given as $\hat{d}_k=\sum_{d=1}^{\bar{N}}\sum_{i=1}^{N_s^d}\bar{w}_k^{d,i}d$.
\end{corollary}
The steps to approximate the posterior densities of state, $\hat{p}(x_k|y_{1:k})$, and delay step, $\hat{p}(j_k|y_{1:k})$ are outlined in Appendix~\ref{algorithm}. Note that when no measurement is received at the estimator, the posterior is approximated with $w_k^{j_k,i}=w_{k-1}^{j_{k-1},i}$ and there is no estimate for delay step.
\section{Simulation Results}
To validate the proposed Gaussian-Approximated filter (GAF) and SMC method for the randomly delayed measurements, we have simulated the two nonlinear state estimation problems: (i) non-stationary growth model and (ii) maneuvering target with unknown and coordinated turn rate. To demonstrate the superiority of the proposed Bayesian estimators, their performances are compared with that of the existing filters for the above two problems. The cubature quadrature sampling points \cite{bhaumik2013cubature} are used to implement the proposed GAF. The particle filter for randomly delayed measurements (PF-RD) developed in \cite{huang2015particle,tiwari2020particle} are reformulated for the proposed measurement model and, along with the standard PF \cite{arulampalam2002tutorial}, are used as the existing filters for performance comparison. The root mean square error (RMSE) \cite{haykin2009cubature} is selected as the performance index for all the implemented filters.   Further, since every particle is assigned a delay at each step, the SMC method is also used to estimate the delay steps, which is not possible with the other implemented filters. 
\subsection{Problem~1}
The time-varying growth model is widely used in literature, owing to its non-stationary property, to validate a newly developed filtering algorithm \cite{hermoso2009unscented,zhang2016particle,arulampalam2002tutorial}. The system model is given as
\begin{equation}\label{non-stationary}
	\begin{split}
		x_k=0.5x_{k-1} + &25\dfrac{x_{k-1}}{1+x_{k-1}^2} + 8 \cos(1.2k)+q_{k-1},\\& z_k={x_k^2}/{20} +v_k,
	\end{split}
\end{equation}
where $q_{k-1}$ and $v_k$ are independent zero mean Gaussian processes with $\mathbb{E}[q_k^2]=10$ and $\mathbb{E}[v_k^2]=1$, respectively. The initial estimate is given by $p(\hat{x}_0) \sim \mathcal{N}(0,1)$ and the number of particles used for the simulation is, $N_s=500$. 
The delayed measurements are generated using a stationary $\lambda_k=\lambda=0.80$ and with maximum permissible delay steps, $N=3$. 
\begin{figure}%
	\centering
	\begin{subfigure}{.49\columnwidth}
		\includegraphics[width=\columnwidth]{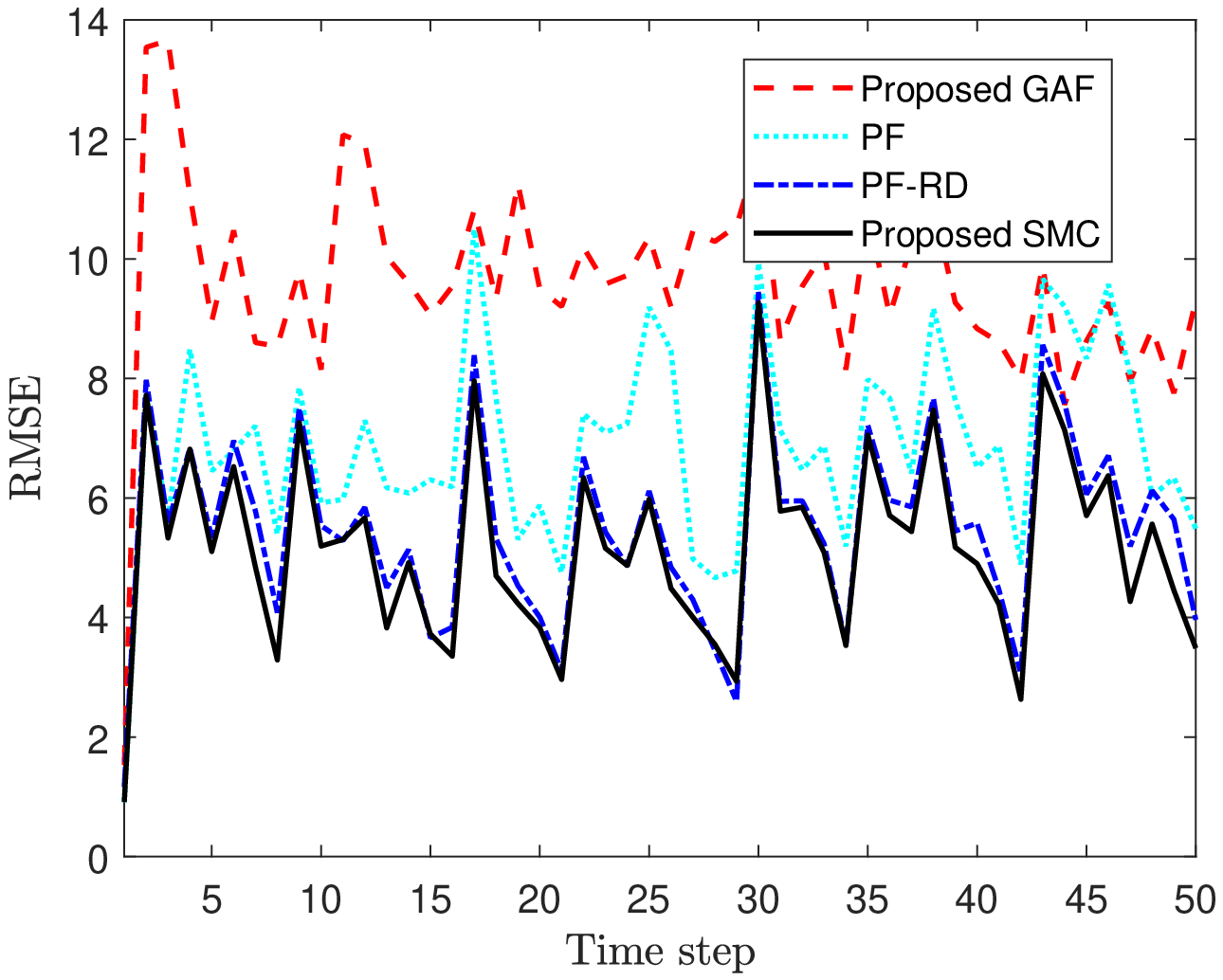}%
		\caption{}%
		\label{2a}%
	\end{subfigure}\hfill%
	\begin{subfigure}{.49\columnwidth}
		\includegraphics[width=\columnwidth]{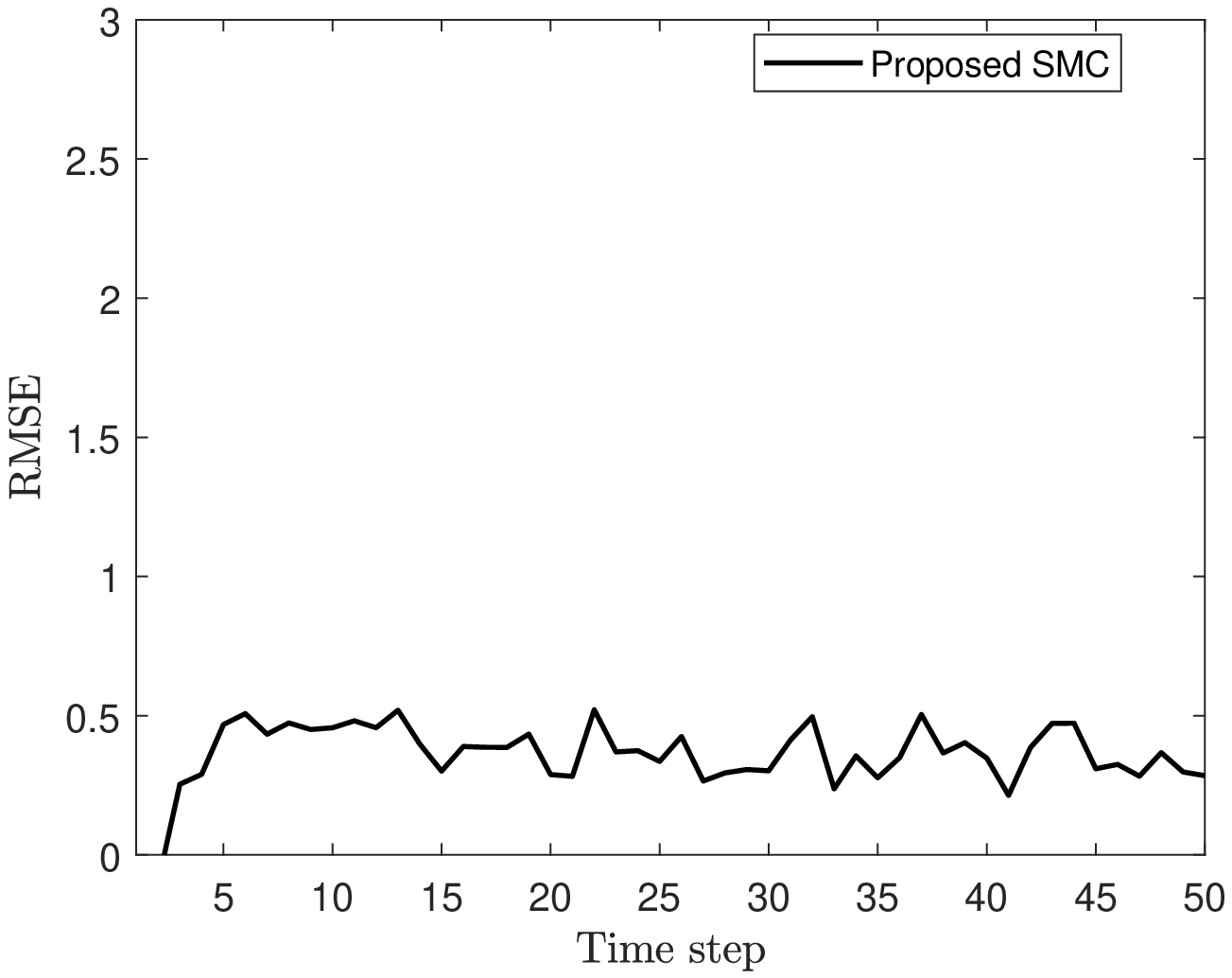}%
		\caption{}%
		\label{2b}%
	\end{subfigure}%
	\caption{(a). RMSE of estimated state. (b). RMSE of estimated delay steps with $N=3$.}
	\label{fig:2}
\end{figure}

To compare the performances, the RMSEs in estimated state calculated by using 100 Monte Carlo (MC) runs are plotted over 50 time steps for each filter in Fig.~\ref{2a}. The time-averaged RMSEs for the proposed SMC, PF-RD, standard PF and the proposed GAF are $5.14$, $5.48$, $7.05$, and $9.60$, receptively. It can be seen that the delay-accounted SMC method and PF-RD perform with more accuracy than the other filters at the cost of additional computational burden which is shown in Table~\ref{rel_time_1}. Fig.~\ref{2b} shows the RMSE in estimated delay using 100 MC runs for the proposed SMC method.
\begin{table}[h]
	\caption{Relative computational time}
		\label{rel_time_1}
	\begin{center}
		\begin{tabular}{|c|c|}
			\hline 
			Algorithms & Relative computational time\\
			\hline \hline
			Proposed GAF & $0.09$\\
			\hline
			Standard PF & $1$  \\
			\hline
			PF-RD & $2$\\
			\hline
			Proposed SMC & $1.75$\\
			\hline
		\end{tabular}
	\end{center}
\end{table}
\subsection{Problem~2}
An aircraft that executes the maneuvering turn in two-dimensional plane with a fixed but unknown turn rate, $\Omega$, is considered by using the coordinated turn model for the aerospace target tracking. This model receives the bearing and range measurement observed from a radar to estimate the unobserved kinematics of the aircraft. The states representing the kinematics of aircraft are $x_k=[\zeta\quad \dot{\zeta}\quad \eta\quad \dot{\eta}\quad \Omega ]^\top$, where $\zeta$ and $\eta$ represent positions, and $\dot{\zeta}$ and $\dot{\eta}$ are used for velocities along the $X$ and $Y$ axes, respectively. The dynamics of the target aircraft in discrete-time is given by \cite{haykin2009cubature,tiwari2020particle}:
\begin{equation}
	\begin{split}
		x_k&=\begin{bmatrix}
			1 & \dfrac{\sin \Omega T}{\Omega} & 0 & -\left(\dfrac{1-\cos \Omega T}{\Omega}\right) & 0\\
			0 & \cos \Omega T & 0 & -\sin \Omega T & 0\\
			0 & \dfrac{1-\cos \Omega T}{\Omega} & 1 & \dfrac{\sin \Omega T}{\Omega} & 0\\
			0 & \sin \Omega T & 0 & \cos \Omega T & 0\\
			0 & 0 & 0 & 0 & 1 \end{bmatrix}x_{k-1}\\&\quad + q_{k-1},
	\end{split}
\end{equation} 
where $T$ is the time interval between two successively received measurements. $q_{k-1}$ is a zero mean Gaussian sequence with covariance $Q=\text{diag}[\text{q}_1M\ \text{q}_1M\ \text{q}_2T]$, where $\text{q}_1$ and $\text{q}_2$ are the noise intensity parameters, and $ M=\begin{bmatrix}
	\frac{T^3}{3} & \frac{T^2}{2}\\
	\frac{T^3}{2} & T \end{bmatrix}$.
The range, $r$, and bearing, $\theta$ are the observation available for tracking, which are observed through a radar placed at the origin. The measurement model can be given as
\begin{equation}\label{eq:48}
	z_k=\big[
	r_k \ \theta_k
	\big]^\top=\Big[ \sqrt{\zeta_k^2+\eta_k^2} \quad \tan^{-1}\left(\dfrac{\eta_k}{\zeta_k}\right)\Big]^\top + v_k,
\end{equation}
where $v_k$ is an independently distributed zero-mean Gaussian sequence with covariance $R=\text{diag}[\sigma_r^2 \quad \sigma_{\theta}^2]$. The parameters used in this simulation are given in Table~\ref{Tab:3}. The initial estimate for state are drawn from the normal distribution with mean and covariance, $\hat{x}_0=[1000$m$\; 300\,\text{m}\text{s}^{-1}\; 1000$m$\; 0\,\text{ms}^{-1}\; -3^o\text{s}^{-1}]^T$ and $\mathbf{P}_{0|0}=\text{diag}[100\text{m}^2\; 10\,\text{m}^2\text{s}^{-2}\; 100\text{m}^2\; 10\text{m}^2\text{s}^{-2}\; 100\text{mrad}^2\text{s}^{-2}]$, respectively. The number of particles used for simulation is, $N_s=5000$. 
The delayed measurements are generated using a stationary $\lambda_k=\lambda=0.90$ and $N=3$.
\begin{table}[h]
	\caption{Tracking parameters}
	\label{Tab:3}
	\begin{center}
		\begin{tabular}{|c|c|}
			\hline
			Sampling Time ($T$) & $0.125$ s\\
			\hline
			Turn Rate ($\Omega$) & $-3^o\text{s}^{-1}$\\
			\hline
			$\text{q}_1$ & $0.1\,\text{m}^2\text{s}^{-3}$ \\
			\hline
			$\text{q}_2$ & $1.75 \times 10^{-4}\text{s}^{-3}$\\
			\hline
			$\sigma_r$ & $10$ m\\
			\hline
			$\sigma_{\theta}$ & $\sqrt{10}$ mrad \\
			\hline
		\end{tabular}
	\end{center}
\end{table} 

\begin{figure}%
	\centering
	\begin{subfigure}{.49\columnwidth}
		\includegraphics[width=\columnwidth]{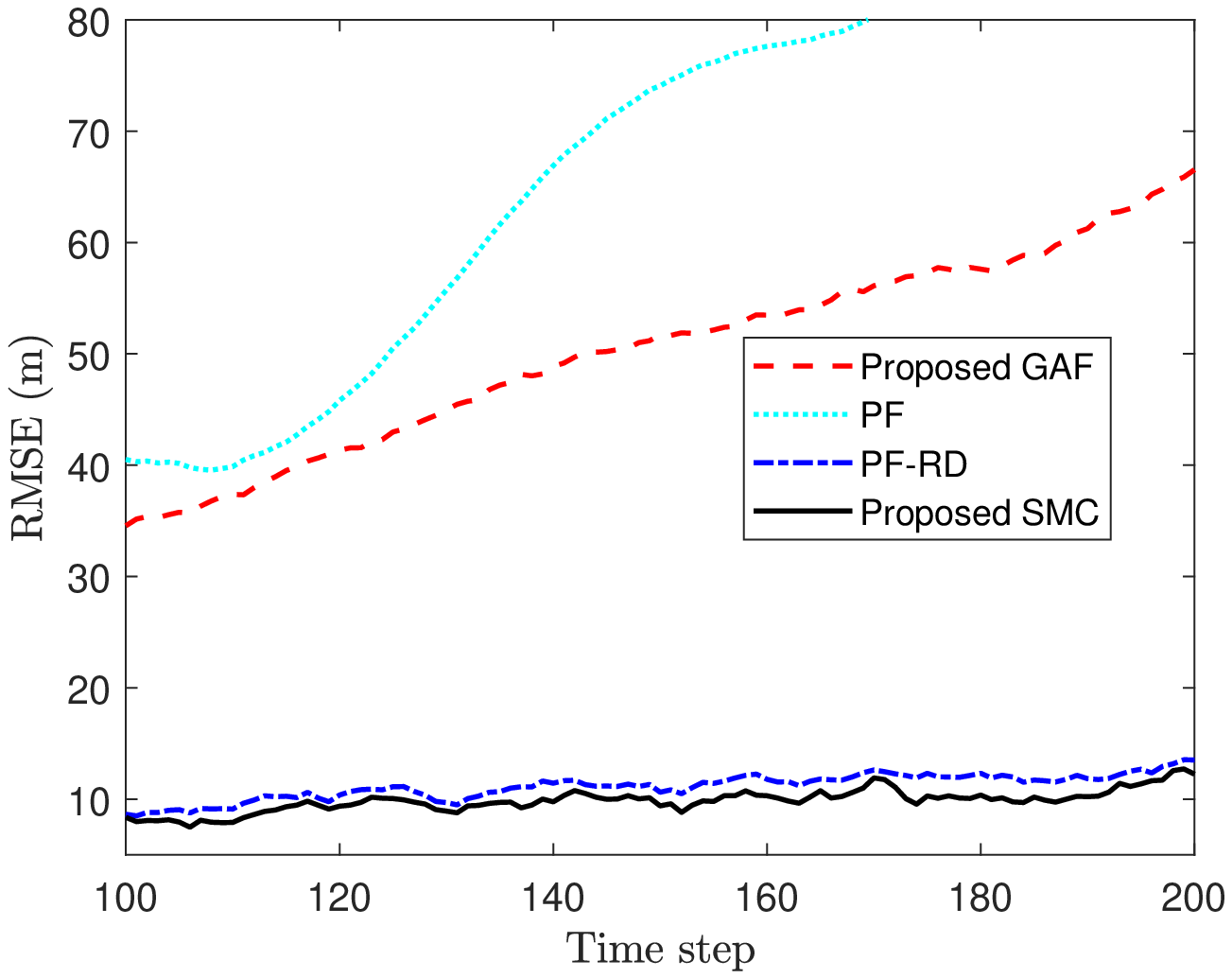}%
		\caption{}%
		\label{3a}%
	\end{subfigure}\hfill%
	\begin{subfigure}{.49\columnwidth}
		\includegraphics[width=\columnwidth]{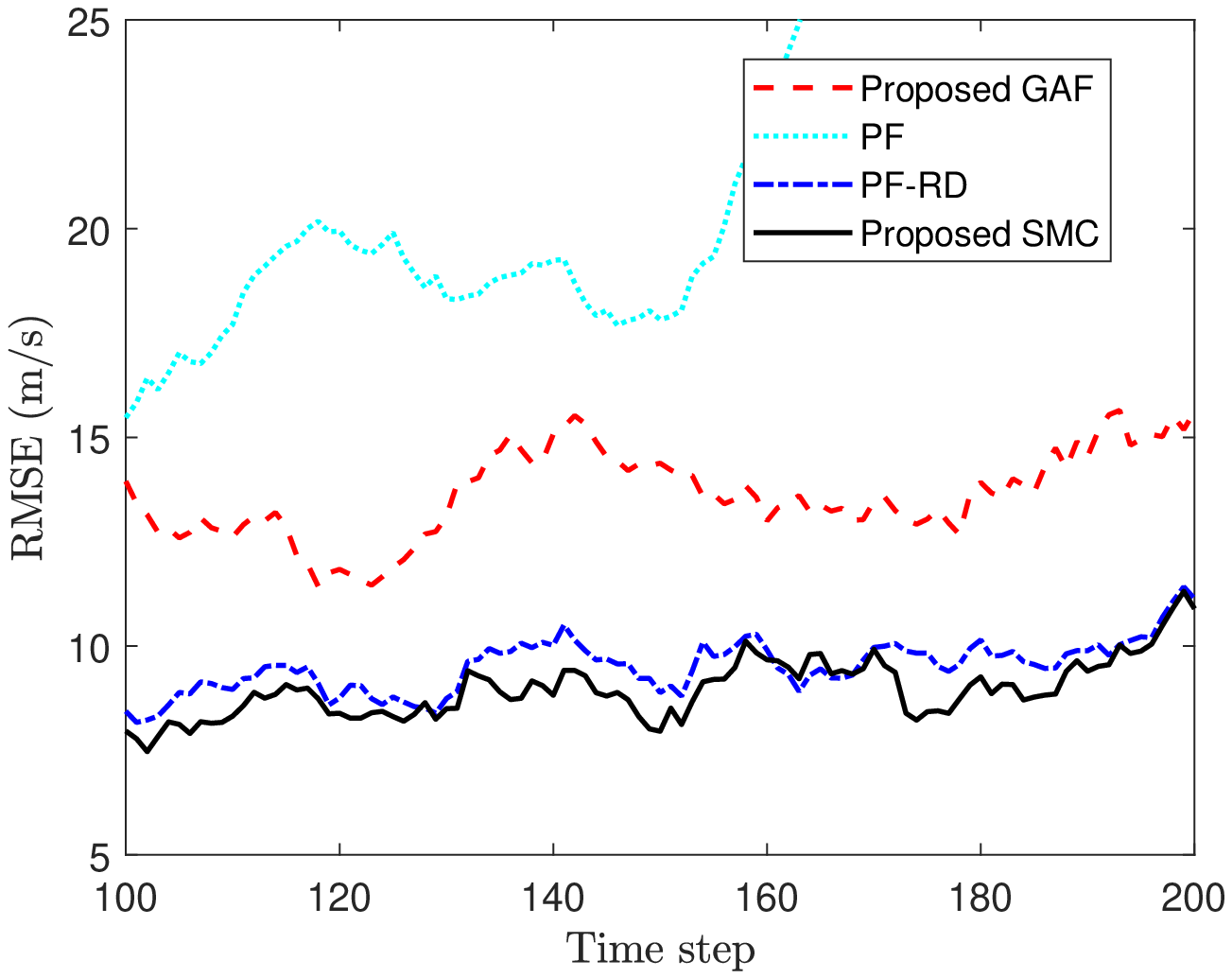}%
		\caption{}%
		\label{3b}%
	\end{subfigure}%
	\caption{(a). RMSE of position. (b). RMSE of velocity.}
	\label{fig:3}
\end{figure}

\begin{figure}%
	\centering
	\begin{subfigure}{.49\columnwidth}
		\includegraphics[width=\columnwidth]{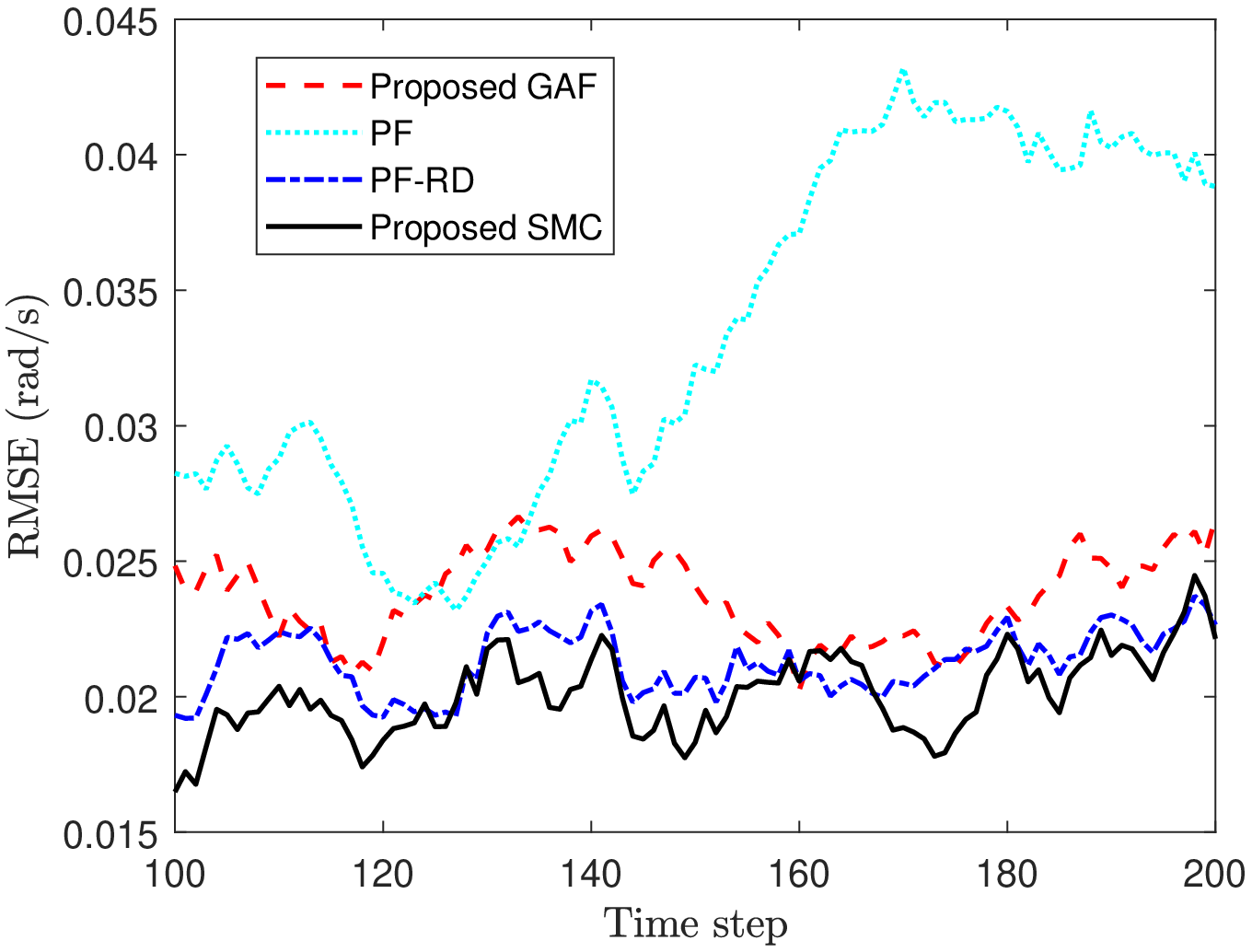}%
		\caption{}%
		\label{4a}%
	\end{subfigure}\hfill%
	\begin{subfigure}{.49\columnwidth}
		\includegraphics[width=\columnwidth]{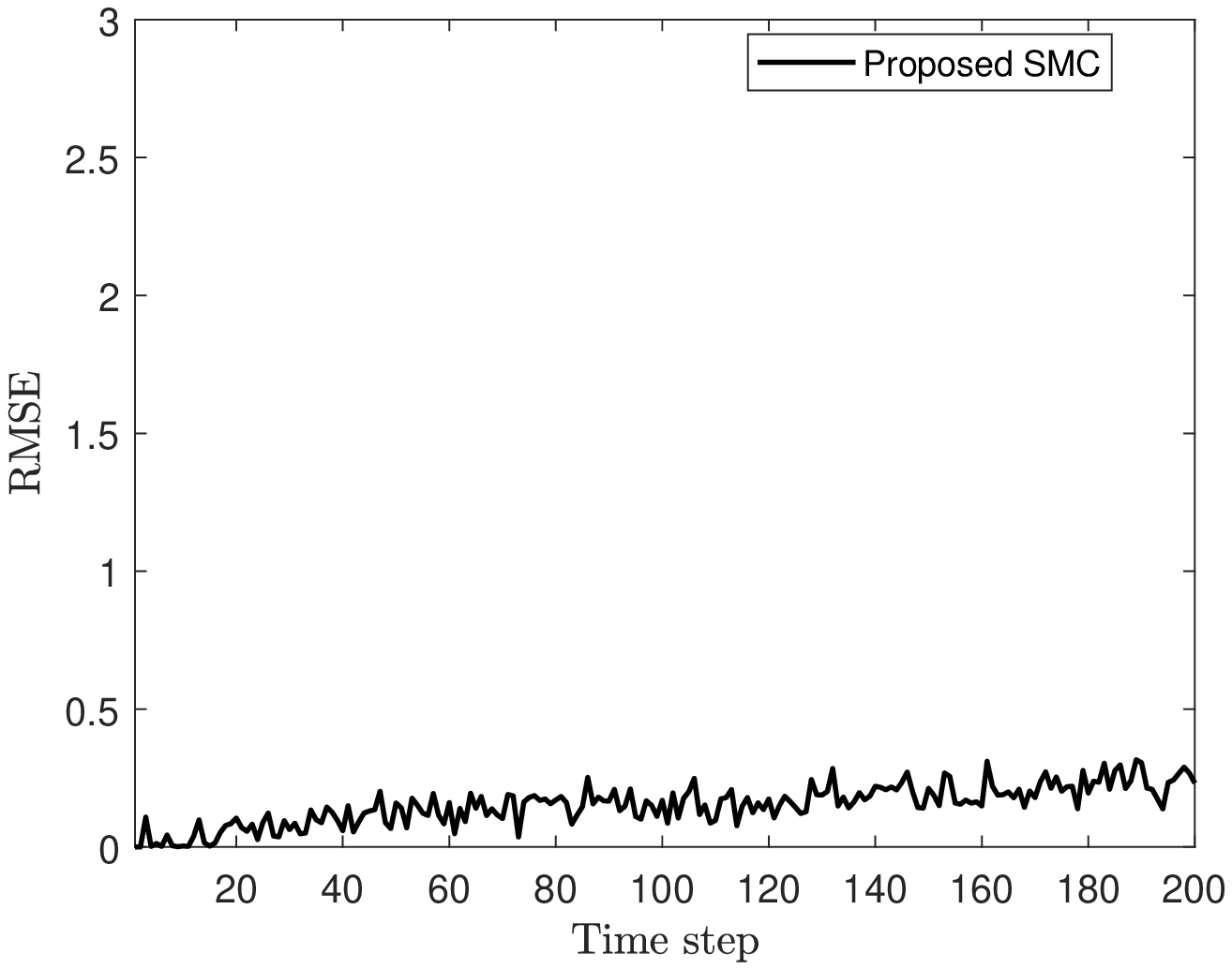}%
		\caption{}%
		\label{4b}%
	\end{subfigure}%
	\caption{(a). RMSE of turn rate. (b). RMSE of delay steps with $N=3$.}
	\label{fig:4}
\end{figure}

The RMSEs calculated over 100 MC runs are plotted in Figs.~\ref{fig:3} and \ref{fig:4} for different filters. It can be observed from the plots that the standard PF, which does not account for the random delays, diverges, whereas the algorithms such as the proposed SMC, PF-RD, and proposed GAF, which have been developed for the random delay, perform with better accuracy. Also, the PF-based filters outperform the GAF at the cost of extra computational effort that is shown in Table~\ref{rel_time_2}. Note that since the particles are not repetitively used to compute the likelihood at each step in the proposed SMC method, it tracks the kinematics of the target with slightly better accuracy than that of the PF-RD at a relatively low computational cost.
 
\begin{table}[h]
	\caption{Relative computational time}
	\label{rel_time_2}
	\begin{center}
		\begin{tabular}{|c|c|}
			\hline 
			Algorithms & Relative computational time\\
			\hline \hline
			Proposed GAF & $0.03$\\
			\hline
			Standard PF & $1$  \\
			\hline
			PF-RD & $2.16$\\
			\hline
			Proposed SMC & $1.26$\\
			\hline
		\end{tabular}
	\end{center}
\end{table}
\section{Conclusions}
This paper presents a measurement model using the Poisson random variable to represent the random delay and packet dropout while receiving the measurements in the networked systems. The proposed model generates the independent measurements and uncorrelated noise sequence over time. Subsequently, we present the generalized Gaussian-approximated filter for the developed delay model by deriving the terms that get modified owing to random delay in measurements. Further, we propose a SMC algorithm for randomly delayed measurements and packet dropouts. This method divides the whole set of samples into several groups based on the possible delay steps. Each sample is assigned with a delay value that represents the number of steps through which measurements are delayed; the delay steps are updated at every time step. The proposed algorithm also gives a method to estimate the delay steps of the received measurement at a time step. The simulation results show that if the filtering algorithms are accounted for random delays, they perform with more accuracy than the conventional filters. Also, the RMSE plots reflect the superiority of the SMC method that is obtained at the cost of an additional computational burden.   
\appendices
\section{Computation of Autocorrelation of Modified Measurement Noise}\label{uncorrelated_noise}
The proposed measurement model in \eqref{delay_model}, by using \eqref{ideal_meas}, can be rewritten as
\begin{equation*}
	y_k=\sum_{j_k=0}^N\beta_k^{j_k} h_{k-j_k}(x_{k-j_k}) + \sum_{j_k=0}^N \beta_k^{j_k}v_{k-j_k},
\end{equation*}
where the modified noise is defined as
\begin{equation}\label{mod_noise}
	\nu_k= \sum_{j_k=0}^N\beta_k^{j_k} v_{k-j_k}.
\end{equation}
Consider that the non-delayed measurement noise has the following property: $\mathbb{E}[v_k]=0$ and $\mathbb{E}[v_kv_k^\top]=R_k,\ \forall k$. Now, using the definition of modified measurement noise, $\nu_k$, in \eqref{mod_noise}, we can establish that $\mathbb{E}[\nu_k]=0$. Further, its autocorrelation can be given as $\mathbb{E}[\nu_k \nu_m^\top]; \ k-m = a,$ where $a$ is any integer number. This can further be computed as
\begin{equation}\label{auto-coor}
	\begin{split}
	\mathbb{E}[\nu_k \nu_m^\top]&=\mathbb{E}\Big[\sum_{s=0}^N \alpha_k^s(1-\alpha_{k-1}^{s-1})\cdots (1-\alpha_{k-s}^0) v_{k-s}\\
	&\quad \times\sum_{l=0}^N\alpha_m^l(1-\alpha_{m-1}^{l-1})\cdots (1-\alpha_{m-l}^0)v_{m-l}^\top\Big]\\
	&=\sum_{s=0}^N\sum_{l=0}^N \mathbb{E}[\alpha_k^s(1-\alpha_{k-1}^{s-1})\cdots (1-\alpha_{k-s}^0)\alpha_{k-a}^l\\&\quad \times(1-\alpha_{k-a-1}^{l-1})\cdots (1-\alpha_{k-a-l}^0)]\mathbb{E}[v_{k-s}v_{k-a-l}^\top].
	\end{split}
\end{equation}
Note that in the above expansion, if $s=0$, there will be only one term, i.e. $\alpha_k^0$, and if $s=1$, the expansion will include $\alpha_{k}^1(1-\alpha_{k-1}^0)$, and so on. The same is true for $l$. Now, considering the different integer values for $a$, \eqref{auto-coor} can be calculated as follows.\\
\underline{\emph{Case-I}. \quad $0<a\leq N$}:\\
It is clear that in above equation\\
$\mathbb{E}[v_{k-s}v_{k-a-l}^\top]=\begin{cases}\mathbb{E}[v_{k-s}v_{k-s}^\top]=R_{k-s};\ \text{if}\ s-l=a\\
	0;\ \text{else}, \end{cases}$
which effectively means that we have to compute the expectation of the first part in \eqref{auto-coor} only for the values of $s$ and  $l$ such that $s-l=a$. The first expectation operation of \eqref{auto-coor} can be expanded as
\begin{equation*}
	\begin{split}
	\sum_{s=0}^N\sum_{l=0}^N& \mathbb{E}[\alpha_k^s(1-\alpha_{k-1}^{s-1})\cdots(1-\alpha_{k-a}^{s-a})\cdots  (1-\alpha_{k-s}^0)\\&\quad \times\alpha_{k-a}^l(1-\alpha_{k-a-1}^{l-1})\cdots (1-\alpha_{k-a-l}^0)],
	\end{split}
\end{equation*}
and for every combination of $s$ and $l$ that gives $s-l=a$, we have a situation where $$\mathbb{E}[(1-\alpha_{k-a}^{s-a})\alpha_{k-a}^l]=\mathbb{E}[(1-\alpha_{k-a}^{l})\alpha_{k-a}^l].$$ Further, given that $\alpha_{k-a}^l$ is a binary variable, we can write $\mathbb{E}[(\alpha_{k-a}^i)^2]=\mathbb{E}[\alpha_{k-a}^l]$, $\mathbb{E}[(1-\alpha_{k-a}^{l})\alpha_{k-a}^l]=0$. Therefore,
$$\mathbb{E}[\nu_k \nu_{k-a}^\top]=0; \quad 0<a\leq N.$$\\
\underline{\emph{Case-II}. \quad $a> N$:}\\
Given that $0\leq s,l\leq N$ and $a>N$, the second expectation of \eqref{auto-coor}, $\mathbb{E}[v_{k-s}v_{k-a-l}^\top]=0, \ \forall s, l.$ Therefore, $$\mathbb{E}[\nu_k \nu_{k-a}^\top]=0; \quad a> N.$$\\
\underline{\emph{Case-III}. \quad $a=0$:}\\
Eq.~\eqref{auto-coor} can be written as
\begin{equation}\label{a=0}
	\begin{split}
		\mathbb{E}[\nu_k \nu_m^\top]&=\sum_{s=0}^N\sum_{l=0}^N \mathbb{E}[\alpha_k^s(1-\alpha_{k-1}^{s-1})\cdots (1-\alpha_{k-s}^0)\alpha_{k}^l\\&\quad \times(1-\alpha_{k-1}^{l-1})\cdots (1-\alpha_{k-l}^0)]\mathbb{E}[v_{k-s}v_{k-l}^\top].
	\end{split}
\end{equation}
Clearly, $$\mathbb{E}[v_{k-s}v_{k-l}^\top]=\begin{cases} \mathbb{E}[v_{k-t}v_{k-t}^\top]=R_{k-t}; \ \text{if}\ s=l=t,\\
	=0;\ \text{else}.\end{cases}$$
Therefore, evaluating \eqref{a=0} for the values of $s$ and $l$ when the both are equal, $s=l=t$, we have
\begin{equation*}
	\begin{split}
			\mathbb{E}[\nu_k \nu_m^\top]&=\sum_{t=0}^N\mathbb{E}[(\alpha_k^t)^2(1-\alpha_{k-1}^{t-1})^2\cdots (1-\alpha_{k-t}^0)^2\\&\quad \times\mathbb{E}[v_{k-t}v_{k-t}^\top].
	\end{split}
\end{equation*}
Since $\alpha_k^t$ $(k=0, 1, \cdots \text{and}\ 0\leq t \leq N)$ are independent and $\mathbb{E}[(1-\alpha_k^t)^2]=\mathbb{E}[(1-\alpha_k^t)]$, we can simplify the above expression as
\begin{equation*}
	\begin{split}
		\mathbb{E}[\nu_k \nu_m^\top]&=\sum_{t=0}^N\mathbb{E}[\alpha_k^t]\mathbb{E}[(1-\alpha_{k-1}^{t-1})]\cdots \mathbb{E}[(1-\alpha_{k-t}^0)]\\&\quad \times\mathbb{E}[v_{k-t}v_{k-t}^\top]\\
		&=\sum_{t=0}^N \gamma_k^t R_{k-t}
	\end{split}
\end{equation*}
\underline{\emph{Case-IV}. \quad $a<0$:}\\
Proceeding similar to Cases-I and II (Cases where, $a>0$), we get $	\mathbb{E}[\nu_k \nu_m^\top]=0;\ a<0.$

Hence, the modified measurement noise, $\nu_k$, in \eqref{mod_noise}, has the following property:
$$\mathbb{E}[\nu_k]=0\ \text{and}\ \mathbb{E}[\nu_k \nu_m^\top]=\begin{cases} \sum_{t=0}^N \gamma_k^t R_{k-t};\ & \text{if}\ k=m\\
	0;\ & \text{if}\  |k-m|>0. \end{cases}$$
\section{Steps for proposed SMC}\label{algorithm}
	\begin{center}
		$[\{\{x_k^{j_k,i},w_k^{j_k,i}\}_{i=1}^{N_s^{j_k}}\}_{j_k=0}^{\bar{N}}]:=\texttt{PROPOSED\_SMC}[\{\{x_{k-1}^{j_{k-1},i},w_{k-1}^{j_{k-1},i}\}_{i=1}^{N_s^{j_{k-1}}}\}_{j_{k-1}=0}^{\bar{N}},\lambda_k,y_k]$
	\end{center}
	\begin{itemize}
		\item \texttt{Set $\bar{N}=\ \texttt{min} (N,k-1)$, and} $\bar\gamma_{k}^{j_k}=\dfrac{\gamma_k^{j_k}}{\sum_{j_k=0}^{\bar{N}}\gamma_k^{j_k}}$.
		\item \texttt{Calculate $N_s^{j_{k-1}}=\bar{\gamma}_{k-1}^{j_{k-1}}$}.
		\item \textit{for} $j_{k-1}=0:\bar{N}$
		\begin{itemize}
			\item \textit{for} $i=1:N_s^{j_{k-1}}$
			\begin{itemize}
				\item[-] \texttt{Draw} $x_k^{j_{k-1},i} \sim q(x_k|x_{0:k-1}^{j_{0:k-1},i}, y_{1:k})$
				\item[-] \texttt{Update the delay of particle by assigning $j_k$ step with probability} 
				$$\gamma_k^{j_k,i}=\begin{cases}
					0;\ \text{if}\  j_k=j_{k-\tau}+\tau, \forall \ 0< \tau \leq \bar{N}\\
					\bar{\gamma}_k^{j_k}; \ \text{else}.
				\end{cases}$$
				\item[-] \texttt{Denote the $j_k$ delay steps assigned to $i$th particle as} $j_{k,i}$. 
				\item[-] \texttt{Evaluate the importance weight with $j_k$ delay as}
				$$w_k^{j_k,i}=w_{k-1}^{j_{k-1},i}p(y_{k}|x_{k-j_k}^{j_k,i})\dfrac{p(x_k^{j_k,i}|x_{k-1}^{j_{k-1},i})}{q(x_k^{j_k,i}|x_{0:k-1}^{j_{0:k-1},i},y_{1:k})}.$$
			\end{itemize}
			\item \textit{end for}
		\end{itemize}
		\item \textit{end for}
		\item \texttt{Normalize the importance weight of the particles as}
		$\bar{w}_k^{j_k,i}=\dfrac{{w}_k^{j_k,i}}{\sum_{j_k=0}^{\bar{N}}\sum_{i=1}^{N_s^{j_k}}{w}_k^{j_k,i}}.$
		\item \texttt{Construct the posterior, $\hat{p}(x_k|y_{1:k})$, as given in} \eqref{particle_approx}.
		\item \texttt{Estimate the delay step as}\\ $\hat{j}_k=\sum_{j_k=1}^{\bar{N}}\sum_{i=1}^{N_s^{j_k}}\bar{w}_k^{j_k,i}j_{k,i}$
		\item \texttt{Resample the particles}
		$[\{\{x_k^{j_k,i},w_k^{j_k,i}\}_{i=1}^{N_s^{j_k}}\}_{j_k=0}^{\bar{N}}]\\:=\texttt{RESAMPLE}[\{\{x_k^{j_k,i},w_k^{j_k,i}\}_{i=1}^{N_s^{j_k}}\}_{j_k=0}^{\bar{N}}]$
	\end{itemize}
%
%
%
%
%
%
\ifCLASSOPTIONcaptionsoff
  \newpage
\fi
%
%
%
%
%
\bibliographystyle{IEEEtran}
\bibliography{IEEEabrv,reference}
\end{document}